\documentclass{article}
\usepackage[affil-it]{authblk}

\usepackage{listings}
\usepackage{makeidx}

\usepackage{graphicx, mathtools, amssymb, mathrsfs}
\usepackage{epstopdf} 
\usepackage{url}
\usepackage{amsmath}
\usepackage{graphicx}
\usepackage{color}
\usepackage[makeroom]{cancel}

\usepackage{float}
\usepackage{cite}   

\usepackage[margin=1in]{geometry}

\newtheorem{theorem}{Theorem}[section]

\newenvironment{definition}[1][Definition]{\begin{trivlist}
\item[\hskip \labelsep {\bfseries #1}]}{\end{trivlist}}

\newcommand{\qed}{\nobreak \ifvmode \relax \else
      \ifdim\lastskip<1.5em \hskip-\lastskip
      \hskip1.5em plus0em minus0.5em \fi \nobreak
      \vrule height0.75em width0.5em depth0.25em\fi}

\let\oldsqrt\sqrt
\def\sqrt{\mathpalette\DHLhksqrt}
\def\DHLhksqrt#1#2{%
\setbox0=\hbox{$#1\oldsqrt{#2\,}$}\dimen0=\ht0
\advance\dimen0-0.2\ht0
\setbox2=\hbox{\vrule height\ht0 depth -\dimen0}%
{\box0\lower0.4pt\box2}}

\DeclareMathOperator{\Tr}{Tr}
\newcommand{\la}{\langle}
\newcommand{\ra}{\rangle}

\begin{document}

\title{Zermelo Navigation in the Quantum Brachistochrone}
\author{Benjamin Russell, Susan Stepney}
\affil{Department of Computer Science, University of York, UK, Y010 5DD}
\date{\today}

\maketitle

\begin{abstract}
We analyse the optimal times for implementing unitary quantum gates in a constrained finite dimensional controlled quantum system.
The family of constraints studied is that the permitted set of (time dependent) Hamiltonians is the unit ball of a norm induced by an inner product on $\mathfrak{su}(n)$.
We also consider a generalisation of this to arbitrary norms.
We construct a Randers metric, by applying a theorem of Shen on Zermelo navigation, the geodesics of which are the time optimal trajectories compatible with the prescribed constraint.
We determine all geodesics and the corresponding time optimal Hamiltonian for a specific constraint on the control i.e. $\kappa\Tr(\hat{H}_c(t)^2) = 1$ for any given value of $\kappa >0$.
Some of the results of Carlini et. al. are re-derived using alternative methods.
A first order system of differential equations for the optimal Hamiltonian is obtained and shown to be of the form of the Euler Poincar\'e equations.
We illustrate that this method can form a methodology for determining which physical substrates are effective at supporting the implementation of fast quantum computation.
\end{abstract}


\section{Problem and Motivation}

\subsection{Implementation of Quantum Gates in Constrained Quantum Systems}

We study the speed that a quantum system can implement a desired quantum gate.
Here, all systems have pure states and finite dimensional Hilbert spaces associated to them.

This question has been discussed from many perspectives before, for example \cite{QCOCU,QOC, qconlan, TFS, Li, AK, KN, Lee, ACAR}.
Notable recent works based on geometric methods are \cite{ACAR, ACAR2, rage}.
Our previous work \cite{mememe} begins an investigation into the application of ``Zermelo Navigation'' to determining speed limits for implementing quantum gates in systems of the form eqn.(\ref{conprb}), by a applying a solution in time optimal control based on Randers geometry \cite{ran}.
We attempt to solve essentially the same problem as Carlini {\it et al} \cite{ACAR2}, but in a way based on intrinsic geometric structures in order to derive a first order equation for the optimal Hamiltonian driving the time evolution operator.

Caneva {\it et al} \cite{OCQSL} address the behavior of a commonly applied numerical algorithm, the Krotov method, near the quantum speed limit \cite{OCQSL}.
Nielsen \cite{NFINCIR} has highlighted a connection between Finlser geometry and quantum optimal control.
Furthermore, this work indicates an interesting connection between quantum circuit complexity and Finsler geometry.
A fuller bibliography on the quantum speed limit more generally can be found in the introduction to \cite{mememe}.

In order to implement a certain quantum information processing (QIP) task in a controlled quantum system, 
we consider the dynamics of the system (more precisely, the time evolution operator $\hat{U}_t$) 
as given by the Schr\"odinger equation:
\begin{align}
\label{conprb}
 \frac{d \hat{U}_t}{dt} = -i\hat{H}_t \hat{U}_t = -i\left(\hat{H}_0 + \hat{H}_c(t)\right)\hat{U}_t
\end{align}
This is a standard form taken by the Schr\"odinger equation in the case of a quantum optimal control problem \cite{QOC}.
$\hat{H}_0$ is the `drift' Hamiltonian, which represents the dynamics of the system in the absence of control fields;
here we take it to be time independent throughout.
$\hat{H}_c(t)$ is the control Hamiltonian, which represents the effect of control fields on the dynamics of $\hat{U}_t$.
We say that a quantum system implements a gate $\hat{O}$ at time $T$ if $\hat{U}_T = \hat{O}$ \cite{ACAR2, mememe, rage}.

\subsection{Central Results of the Paper}

We show that, under the assumption that $\hat{H}_c(t)$ is constrained such that $h(i\hat{H}_c(t), i\hat{H}_c(t)) = 1$ for some inner product on $\mathfrak{su}(n)$, the time optimal trajectories of the time evolution operator $\hat{U}_t$ are exactly the geodesics of the following right invariant Randers metric on $SU(n)$:
\begin{align}
 F_{\hat{U}}(\hat{A} \hat{U}) & = \sqrt{\frac{h(\hat{A}, \hat{A})}{\lambda} + \frac{h(\hat{A}, i\hat{H}_0)^2}{\lambda^2}} + \frac{h(\hat{A}, i\hat{H}_0)}{\lambda}
\end{align}
where $\lambda = 1 - h(i\hat{H}_0, i\hat{H}_0)$.

We show that, under the assumption that $\hat{H}_c(t)$ is constrained such that $\kappa \Tr(\hat{H}_c(t)^2) =1$ (that is, choosing $h$ to be some multiple of the Killing form of $\mathfrak{su}(n)$), the time optimal trajectories are given by:
\begin{align}
\label{geos}
 \hat{V}_t 
	   & = \exp(-it\hat{H}_0) \exp (i t \hat{D}) \nonumber
\end{align}
for some $i\hat{D}$ (depending on the desired gate) in $\mathfrak{su}(n)$ such that $h(i\hat{D},i\hat{D})= \kappa\Tr(\hat{D}^2)=1$.

We show that, under the same constraint, the optimal control Hamiltonian is:
\begin{align}
\hat{H}_c(t) = - \exp(-it\hat{H}_0) \hat{D} \exp(it\hat{H}_0)
\end{align}

We give a series expansion based on the Baker--Campbell--Hausdorff formula for $\hat{D}$ in terms an arbitrary desired gate $\hat{O} \in SU(n)$.

We present a set of equations (\ref{toe}) which determine the time optimal control Hamiltonian for an arbitrary gate in any situation where $\check{F}(i\hat{H}_c(t)) = 1$ for an arbitrary Minkowski norm $\check{F}$ on $\mathfrak{su}(n)$ in scenarios where $\check{F}(i\hat{H}_0) \leq 1$ for that same norm.


\section{Speed Limits in Constrained Systems}

In this section we outline the problem of Zermelo navigation on a manifold.
We illustrate how a Randers metric, which depends on the `drift' dynamics of a controlled dynamical system on a manifold and the driftless speed of the navigator, has the property that its geodesics are time optimal trajectories for navigating on the same manifold.
We discuss Shen's theorem \cite{Shen} which provides an explicit formula for this Randers metric.
We are motivated by the desire to determine time optimal trajectories in quantum systems with constraints in novel ways.
This method builds on previous work as it allows one to obtain the optimal time for \emph{arbitrary} trajectories, as well as determining theoretically optimal ones.
This is practically relevant as it is not always the case that all trajectories can be practically realised.

In the following we use the notation $T_pM$ to refer to the tangent space at the point $p$ on a manifold $M$, and $TM$ to refer to the entire tangent bundle of a manifold, that is, all tangent spaces considered together.
For example $T_{\hat{U}} SU(n)$ is the tangent space to $SU(n)$ at the point $\hat{U} \in SU(n)$.
We use the notation $\Gamma\left(TM\right)$ to refer to the set of all smooth sections of the tangent bundle $TM$,
that is, effectively to say, all smooth vector fields on $M$.
For good references using this notation for both general manifolds and Lie groups, see \cite{Bump, Lee}.

\subsection{Randers Metrics and Shen's Theorem on Zermelo Navigation}

\subsubsection{Definition of a Finsler Metric}

A Finsler metric on a manifold $M$, of which a Randers metric is a special case, is a type of metric structure generalising a Riemannian metric.
This generalisation is essentially that the norm induced on each tangent space does not need to be induced by an inner product.
Technically, a Finsler metric on a smooth manifold $M$ is a smoothly varying `Minkowski norm' on each tangent space on $M$.

\begin{definition}
A \emph{Minkowski Norm} on a vector space $V$ is a function $|\cdot| : V \rightarrow \mathbb{R}$ which satisfies the following axioms:
 \begin{itemize}
  \item $\forall v \in V$, $|v| \geq 0$. $|v| = 0 \Leftrightarrow v=0$. Positive definite.
  \item $\forall v \in V, \forall \lambda \in \mathbb{R}^{+}$, $|\lambda v| = \lambda |v|$. Positive homogeneous.
  \item $\forall u, v \in V, |u + v| \leq |v| + |u|$. Subadditive.
  \item $\forall v, x, y \in V$ the Hessian of $|\cdot|^2$ is positive definite.
  That is: $\frac{1}{2} \left. \frac{\partial^2}{\partial s \partial t} {|v + tx + sy|^2} \right|_{t=0, s=0} \geq 0$
 \end{itemize}
\end{definition}

The only Minkowski norms that are true norms are exactly those that are reversible: $|-v| = |v|$ $\forall v \in V$.
For a detailed discussion of this definition see \cite{Fins, ran}.
See \cite{OTD} for an interesting and thorough discussion of variations of Finsler metrics in use in applications.

\subsubsection{Zermelo Navigation}

We note a type of metric for which the lengths of any curve is the optimal traversal time rather than just a bound, namely Randers metrics.
A Randers metric is a specific case of a Finsler metric \cite{ran}.
A Randers metric on a manifold $M$ is a Finsler metric on $M$ which is a Randers norm on each tangent space.

\begin{definition}
 A \emph{Randers} norm on a real, finite dimensional vector space $\mathbb{R}^N$ is a map $ |\cdot| : \mathbb{R}^N \rightarrow \mathbb{R}$ which can be written as follows:
 \begin{align}
  | v | = \sqrt{\alpha(v, v)} + \beta(v)
 \end{align}
 where $\alpha$ is an inner product on $\mathbb{R}^N$ and $\beta$ is a one form.
\end{definition}

\begin{definition}
Subsequently, a \emph{Randers metric} on a manifold $M$ can always be written as a Finlser metric with fundamental function $F_p: T_pM \rightarrow \mathbb{R}$ of the following form:
 \begin{align}
  F_p(v) = \sqrt{\alpha_p(v,v)} + \beta_p(v)
 \end{align}
 for any $v \in T_p M$, where $\alpha$ is a Riemannian metric in the usual sense and $\beta$ is a differential one form.
\end{definition}

A theorem due to Shen relates the problem of Zermelo navigation on a Riemannian manifold $(M,h)$ in the presence of vector field $W$ (such that $h_p(W_p, W_p) \leq 1$, $\forall p \in M$) to the geodesics of Randers metrics \cite[Ch.2]{ran}.
The results of Shen \cite{Shen} completely classify Randers metrics and demonstrate that they are in one to one correspondence with Zermelo navigation problems on Riemannian manifolds.

The problem of Zermelo navigation on a Riemannian manifold can be stated as follows \cite{Bao2004}.
Given:
\begin{itemize}
 \item A Riemannian manifold $(M,h)$
 \item A smooth vector field $W \in \Gamma(TM)$ such that $h_p(W_p, W_p) \leq 1$, $\forall p \in M$
\end{itemize}
determine the time optimal trajectories to follow in order to navigate from an arbitrary given initial point $p_I$ to an arbitrary given terminal point $p_T$ on $M$, where the navigator is restricted such that its speed (according to $h$) in the absence of wind (i.e. $W=0$) is exactly $1$. 

Shen's theorem can be stated as follows: the time optimal trajectories are exactly the geodesics of the Randers metric $F$ defined by the following equations:
\begin{align}
\label{thingie}
 F_p(v) & = \sqrt{\alpha_p(v,v)} + \beta_p(v) \\
 \alpha_p(u,v) & = \frac{h_p(u,v) + \lambda_p \beta_p (u)\beta_p(v) }{\lambda_p} \nonumber \\
 \beta_p(v) & = -\frac{h_p(v,W_p)}{\lambda_p} \nonumber \\
 \lambda_p & = 1-h_p(W_p,W_p) \nonumber
\end{align}
where these definitions hold  $\forall u,v \in T_p M, \forall p \in M$.
The optimal times are the geodesic lengths of the same metric.
$(M,h)$ and $W$ together are know as the navigation data for the specific instance of Zermelo navigation.
For a full derivation of these facts and much more detailed discussion providing stronger intuitive aids, see \cite[Ch.2]{ran}.
For a recent application to quantum computing see \cite{mememe}.

\subsection{Right Invariance and Right Extension}
It is well known that $SU(n)$ is a compact, connected Lie group.
Here we state some standard definitions from Lie theory (see for example \cite{Bump, hall,LGBAI}), relevant for our later derivations.

\begin{definition}
A function $F:T SU(n) \rightarrow \mathbb{R}$ is said to be \emph{Right Invariant} if the following holds:
\begin{align}
 F_{R_{\hat{V}} (\hat{U}) } (dR_{\hat{V}} \big|_{\hat{U}} (\hat{A})) = F_{\hat{U}}(\hat{A})
\end{align}
$\forall \hat{U}, \hat{V} \in SU(n), \forall \hat{A} \in T_{\hat{U}} SU(n)$.
As $SU(n)$ is a linear algebraic group (or LAG) \cite{LAGS} this definition simplifies to:
\begin{align}
 F_{\hat{U} \hat{V}} (\hat{A} \hat{V}) = F_{\hat{U}}(\hat{A})
\end{align}
\end{definition}

\begin{definition}
A function $F:T SU(n) \rightarrow \mathbb{R}$ is said to be the \emph{Right Extension} of $G: T_{\hat{I}} SU(n) \rightarrow \mathbb{R}$ if the following holds:
\begin{align}
 F_{\hat{U}}(\hat{A}) = F_{\hat{I}}(\hat{A} \hat{U}^{-1})
\end{align}
\end{definition}
One readily checks that right extension always leads to a right invariant function on $TM$.

This construction allows one to construct all right invariant Randers metrics on $SU(n)$ as they are in one to one correspondence with pairs of the form $(\alpha, \beta)$ wherein $\alpha : \mathfrak{su}(n) \times \mathfrak{su}(n) \rightarrow \mathbb{R}$ is an inner product on $\mathfrak{su}(n)$ (when identified with the tangent space at the identity) and $\beta : \mathfrak{su}(n) \rightarrow \mathbb{R}$ is a one form on $\mathfrak{su}(n)$.
$\alpha$ and $\beta$ can be combined into a Randers norm $F: \mathfrak{su}(n) \rightarrow \mathbb{R}$ by setting $F(\hat{A}) = \sqrt{\alpha(\hat{A}, \hat{A})} + \beta(\hat{A})$.
Now the metric can be set to be the right extension of this function to the entire tangent bundle of $SU(n)$ as follows:
\begin{align}
 F_{\hat{U}}(\hat{A} \hat{U}) = F_{\hat{I}}(\hat{A}) = \sqrt{\alpha(\hat{A}, \hat{A})} + \beta(\hat{A})
\end{align}
One can also easily check that $F$ is a smooth function on $TM/\{0\}$ (the slit tangent bundle) in the required sense to be a Finsler metric.
This follows from the fact that right translation in a Lie group is smooth as the group multiplication operation is smooth.

One could equally right translate $\alpha$ and $\beta$ before adding them, but this would yield the same result.
It it readily checked that a Randers metric is right invariant exactly when its Riemannian and linear parts are right invariant individually.
This fact precludes the possibility of a bi-invariant Randers metric on $SU(n)$ as there are no bi-invariant one forms on $SU(n)$.

\subsection{Shen's Theorem Applied to Right Invariant Metrics on $SU(n)$ and Quantum Mechanics}

We now set up the problem of Zermelo navigation on $SU(n)$ and show how it can be applied to quantum mechanics.
Suppose that $h : \mathfrak{su}(n) \times \mathfrak{su}(n) \rightarrow \mathbb{R}$ is an inner product on $\mathfrak{su}(n)$.
Suppose that a controlled quantum system of the form eqn.(\ref{conprb}) is constrained such that its control Hamiltonian $\hat{H}_c(t)$ satisfies $h(i\hat{H}_{c}(t), i\hat{H}_{c}(t)) = 1$, $\forall t$.
That is, the constraint is time independent and satisfied for all time.
It is clear that the right invariance of the Riemannian metric $h_{\hat{U}}$ (formed by right extending $h$) corresponds to a constraint only on $\hat{H}_t$, not any constraint depending on $\hat{U}_t$ explicitly.
This argument is very similar to the argument for the requirement of right invariance presented in \cite{NFINCIR}.

Suppose further that a drift Hamiltonian $\hat{H}_0$ is given.
The time evolution operator for our quantum system now satisfies a Schr\"odinger equation of the form eqn.(\ref{conprb}).
The tangent vector $\frac{d \hat{U}_t}{dt}$ to the curve $\hat{U}_t$ has two terms: $-i \hat{H}_c(t) \hat{U}_t$ and $-i\hat{H}_0 \hat{U}_t$.
In order to fix terminology closer to the original formulation of the Zermelo navigation problem, we define the ``wind'' vector field on $SU(n)$ by $\hat{W}_{\hat{U}} = -i\hat{H}_0 \hat{U}$.

In such a setup, there is enough information to construct the ``navigation data'' for a Zermelo navigation problem.
From these ingredients one can construct a Finsler metric (which is in fact a Randers metric) that has the property that its geodesics are the time optimal trajectories for $\hat{U}_t$ to be driven between given endpoints, by applying Shen's theorem, eqn.(\ref{thingie}).
This Randers metric is $F_{\hat{U}}(\hat{A}) = \sqrt{\alpha_{\hat{U}}(\hat{A}, \hat{A})} + \beta_{\hat{U}}(\hat{A})$.
In terms of the navigation data $(h, \hat{W})$ on $SU(n)$, the $\alpha$ and $\beta$ are found to be:
\begin{align}
 \alpha_{\hat{U}}(\hat{A}\hat{U}, \hat{A}\hat{U}) & = \frac{\lambda h_{\hat{U}}(\hat{A}\hat{U}, \hat{A}\hat{U}) + h_{\hat{U}}(\hat{A}\hat{U}, \hat{W}_{\hat{U}})^2}{\lambda^2} \\
 & = \frac{h_{\hat{U}}(\hat{A}\hat{U}, \hat{A}\hat{U})}{\lambda} + \frac{h_{\hat{U}}(\hat{A}\hat{U}, -i\hat{H}_0 \hat{U})^2}{\lambda^2} \nonumber \\
 & = \frac{h(\hat{A}, \hat{A})}{\lambda} + \frac{h(\hat{A}, i\hat{H}_0)^2}{\lambda^2} \nonumber
\end{align}
\begin{align}
 \beta_{\hat{U}}(\hat{A}\hat{U}) = \frac{h_{\hat{U}}(\hat{A}\hat{U}, -i\hat{H}_0 \hat{U})}{1- h_{\hat{U}}(-i\hat{H}_0\hat{U}, -i\hat{H}_0\hat{U})} = \frac{h(\hat{A}, -i\hat{H}_0)}{{1- h(i\hat{H}_0, i\hat{H}_0)}} = \frac{h(\hat{A}, -i\hat{H}_0)}{\lambda}
\end{align}
Thus $F$ is, in full, given by:
\begin{align}
 F_{\hat{U}}(\hat{A} \hat{U}) & = \sqrt{\frac{h(\hat{A}, \hat{A})}{\lambda} + \frac{h(\hat{A}, i\hat{H}_0)^2}{\lambda^2}} + \frac{h(\hat{A}, -i\hat{H}_0)}{\lambda} = F(\hat{A})
\end{align}
As stated above, the right invariance of this quantity is clear.
Here $\lambda_{\hat{U}} = 1- h_{\hat{U}}(\hat{A}\hat{U}, \hat{A}\hat{U}) = 1- h(\hat{A}, \hat{A})$, thus we note that $\lambda$ is a scalar quantity because it is right invariant, and that all right invariant scalar quantities are constant.
This is a simplifying factor of the case when $h$ and $W$ are right invariant compared to the general case.

\subsection{The Length Functional when $F$ is Right Invariant}

In Riemannian geometry, each Riemannian metric defines a length functional on the space of all curves on a Riemannian manifold.
The situation in Finsler geometry is essentially identical.

The length functional $L[\hat{U}_t]$ for a curve $\hat{U}_t : [0,T] \rightarrow SU(n)$, for a Finlser metric can be written as follows:
\begin{align}
 \label{winmet}
 L[\hat{U}_t] = \int_{t=0}^{T} F_{\hat{U}_t}\left(\frac{d \hat{U}_t}{dt}\right) dt
\end{align}
In the case that $F$ is is right invariant one finds:
\begin{align}
 L[\hat{U}_t] & = \int_{t=0}^{T} F\left(\frac{d \hat{U}_t}{dt} \hat{U}_{t}^{-1} \right) dt \nonumber
\end{align}
In the case that $\hat{U}_t$ solves the Schr\"odinger equation (\ref{conprb}) one finds:
\begin{align}
 L[\hat{U}_t] & = \int_{t=0}^{T} F_{\hat{U}_t}\left(\frac{d \hat{U}_t}{dt}\right) dt 
    = \int_{t=0}^{T} F_{\hat{U}_t}\left(-i\hat{H}(t) \hat{U}_t \right) dt\\
 & = \int_{t=0}^{T} F\left(-i\hat{H}(t) \right) dt 
    = \int_{t=0}^{T} F\left(-i\hat{H}_0 - i\hat{H}_c(t) \right) dt \nonumber
\end{align}
The length $L[\hat{U}_t]$ depends only on quantities in $\mathfrak{su}(n)$ rather than on the group in general, as all dependence on $\hat{U}_t$ itself has disappeared.
It is possible to formulate the geodesic equation for such a Finsler metric as an ODE in $\mathfrak{su}(n)$,
and leads to an equation for the time optimal control Hamiltonian for a controlled quantum system with the type of constraint discussed above.

\subsection{Geodesics of Right Invariant Randers Metrics on $SU(n)$}

This shows that the desired time optimal trajectories required are the geodesics of a right invariant Randers metric on $SU(n)$.
In order to find these geodesics we must determine the extremal curves of the length functional for a Randers metric:
\begin{align}
 L[\hat{U}_t] = \int_{0}^{T} \left[ \sqrt{\alpha_{\hat{U}_t} \left(\frac{d \hat{U}_t}{dt}, \frac{d \hat{U}_t}{dt}\right)} + \beta_{\hat{U}_t} \left(\frac{d \hat{U}_t}{dt} \right) \right] dt
\end{align}
under the assumption of right invariance.

This could be achieved via the usual Euler--Lagrange (EL) equations.
However another method exists that exploits the right invariance of the metric.
By considering the quantity $\frac{d \hat{U}_t}{dt} \hat{U}_t^{-1} = \frac{d \hat{U}_t}{dt} \hat{U}_t^{\dagger}$, we can determine a first order differential equation for its value along a geodesic, the Euler--Poincar\'e (EP) equations (\cite{lre, Hol}; see \cite{EPop} for an application to optimal control)
by the procedure of Lagrangian reduction.
Furthermore, we have $\frac{d \hat{U}_t}{dt} \hat{U}_t^{\dagger} = -i\hat{H}_t$, by applying the Schr\"odinger equation for any potentially time dependent quantum system.
These yield a first order equation for the Hamiltonian which drives the system along a geodesic of any given Randers metric.

The geodesics of Randers metrics have already been seen to be the time optimal trajectories for the relevant constraint, so the EP equations corresponding to a Randers metric on $SU(n)$ are a first order system of equations for the Hamiltonian driving $\hat{U}_t$ along a time optimal trajectory.
This illustrates an alternative understanding of the origin of the first order ``Brachistochrone'' equation \cite{ACAR2}, at least in a special case.
Applying boundary conditions to such an equation allows us to obtain the Hamiltonian that drives $\hat{U}_t$ from $\hat{I}$ to a desired operator (i.e. a specific quantum gate) in the least time.

In a coordinate-free language (where $\xi \in \mathfrak{su}(n)$) the EP equation reads \cite{lre,Hol}:
\begin{align}
 \frac{d}{dt} \frac{\partial \ell}{\partial \xi} = - \text{ad}^{*}_{\xi}\left(\frac{\partial \ell}{\partial \xi}\right)
\end{align}
where $\ell:\mathfrak{su}(n) \rightarrow \mathbb{R}$ is the restriction of an arbitrary right invariant Lagrangian $\mathcal{L}:TSU(n) \rightarrow \mathbb{R}$ to $\mathfrak{su}(n)$,
and $\text{ad}^{*}$ is the co-adjoint representation of $\mathfrak{su}(n)$ \cite{RLG}.
Note the minus sign here: this is due to the metric here being right invariant rather than left invariant as is more commonly studied in pure mathematics contexts such as \cite{Hol}.

There are some additional conditions on $\mathcal{L}$ for the EP equations to apply;
these can be readily found in any mathematical description of the theory of Lagrangian reduction \cite{EPop,lre}.
It is clear that all Finsler metrics meet the required conditions.
For example, it is clear that the regularity condition is met, as it is present in the definition of a Finsler metric.

This equation may also been seen with a $\delta$ (signifying a functional derivative) in place of the $d$ above; this is the form of the equation which applies to infinite dimensional problems rather than the finite dimensional ones studied here.

On fixing a basis $\{\hat{B}_k \}$ for $\mathfrak{su}(n)$ and expressing an arbitrary element $-i\hat{H}_t$ as $\xi^k \hat{B}_k$, the EP equation takes the form \cite{lre, Hol}:
\begin{align}
\label{somelabel}
 \frac{d}{dt} \frac{\partial \ell} {\partial \xi^{d}} = - C^{b}_{a d} \frac{\partial \ell}{\partial \xi^b} \xi^a
\end{align}
where $C^{d}_{a b}$ are the structure constants of $\mathfrak{su}(n)$.  
See \cite{hall} for details of structure constants in general,
and \cite{plie} for $\mathfrak{su}(n)$ specifically,
where the structure constants of $\mathfrak{su}(2)$ and $\mathfrak{su}(4)$ are given explicitly.
The tensor $C$ possesses many symmetries, including $C^{a}_{bd} = -C^{a}_{db}$ for example; this follows directly from the antisymmetry of the Lie bracket.  The use of these symmetries as a tool for simplifying the EP equations in the case of $\mathfrak{su}(n)$ will be included in further work.

Henceforth the subscripts indicating a point on $SU(n)$ are dropped from $\alpha$ and $\beta$, and they are understood to be restricted to the tangent space of $SU(n)$ at the identity, i.e. $\mathfrak{su}(n)$.
However, coordinate indicies still appear.

Setting $\ell$ to be the square (to obtain unit speed geodesics) Randers norm $\ell(\hat{A}) = \frac{1}{2} (F \big|_{\hat{I}}(\hat{A}))^2 = \frac{1}{2} \left(\sqrt{\alpha(\hat{A}, \hat{A})} + \beta(\hat{A})\right)^2$, i.e. the restriction of a Randers metric $F$ on $SU(n)$ to $\mathfrak{su}(n)$, we can derive the EP equation associated to the geodesics of $F$.

Substituting:
\begin{align}
 \frac{\partial \ell }{\partial \xi^d} = 
         \left( (\alpha_{ij}\xi^i \xi^j)^{1/2} + \beta_k \xi^k \right) 
        \left( ||\xi||_{\alpha}^{-1} \alpha_{n d} \xi^n + \beta_d \right)
\end{align}
Differentiating:
\begin{align}
 \frac{d}{dt} \left(\frac{\partial \ell }{\partial \xi^d}\right) & = 
        \frac{d}{dt} \left( ||\xi||_{\alpha} + \beta_k \xi^k \right) 
        \left( ||\xi||_{\alpha}^{-1} \alpha_{n d} \xi^n + \beta_d\right) \\
 & = \left( ||\xi||_{\alpha}^{-1} \langle \dot{\xi}, \xi \rangle_{\alpha} + \beta_{j}\dot{\xi}^j\right)
        \left( ||\xi||_{\alpha}^{-1}\alpha_{md}\xi^m +\beta_d \right) - \nonumber \\
 & \qquad \left( ||\xi||_{\alpha} + \beta_k\xi^k \right)\left( ||\xi||_{\alpha}^{-3}\langle \xi, \dot{\xi} \rangle_{\alpha} \alpha_{hd}\xi^h - ||\xi||_{\alpha}^{-1} \alpha_{kd}\dot{\xi}^k \right) \nonumber
\end{align}
These yield the EP equation of a geodesic:
\begin{align}
\label{geoeqn1}
\left( ||\xi||_{\alpha}^{-1} \langle \dot{\xi}, \xi \rangle_{\alpha} + \beta_{j}\dot{\xi}^j\right)
    \left( ||\xi||_{\alpha}^{-1}\alpha_{md}\xi^m +\beta_d \right) 
    - \qquad &\\
\left( ||\xi||_{\alpha} + \beta_k \xi^k \right)
    \left( ||\xi||_{\alpha}^{-3} \langle \xi , \dot{\xi} \rangle_{\alpha} \alpha_{hd}\xi^h 
        - ||\xi||_{\alpha}^{-1} \alpha_{kd}\dot{\xi}^k \right) 
    &  \nonumber \\
= -C^{a}_{bd} \left(( ||\xi||_{\alpha} + \beta_k \xi^k ) \left( ||\xi||_{\alpha}^{-1} \alpha_{n a} \xi^n + \beta_a\right)\right) \xi^b \nonumber
\end{align}
where we take the following meanings: $||\xi||_{\alpha} := \sqrt{\alpha_{ij} \xi^i\xi^j}$ and $\langle \mu, \nu \rangle_{\alpha} := \alpha_{ij}\mu^i\nu^j$.

We are interested in the geodesics associated to a navigation problem specified in terms of its navigation data.
Such an equation can be obtained by substituting in the definitions of $\alpha$ and $\beta$ in terms of $h$ and $W$ from Shen's solution to the navigation problem.
This is an elementary but somewhat tedious computation that provides little insight, so is omitted here.
Many numerical/approximate methods exist for solving this type of first order ODE.
Thus, many practical approaches could be taken to solving for the optimal Hamiltonian after obtaining the equation in terms of $h$ and $\hat{O}$.
We leave obtaining this final equation to the reader.

\subsection{Applications of Killing Fields to Determining Geodesics}

Choose $h(\hat{A}, \hat{A}) = \kappa \Tr(\hat{A}^{\dagger} \hat{A})$ (i.e. a constant positive multiple $\kappa$ of the Killing form); then the right extension of $h$ is the unique bi-invariant metric.
This is essentially the case studied in \cite{ACAR2, mememe}.
It exhibits a simplifying factor pertaining to the task of determining geodesics.
The geodesics can be determined by an application of a special case of Robles \cite[thm.2]{CRob}. 
We use $\sigma=0$ in that theorem, as the special case of a Killing field (see \cite{FinNav} for definitions) in place of the infinitesimal homothety.
Also, we specialise to $SU(n)$, rather than a general manifold, as this is the case relevant to quantum mechanics.
In fact, there are no infinitesimal homotheties that are not Killing fields for the bi-invariant metric on $SU(n)$, so no real restriction has been incurred on which metrics can have their geodesics determined using the following theorem.
The theorem states:
\begin{theorem} [adapted from {\cite[thm.2]{CRob}}]
\label{great}
Given:
\begin{itemize}
 \item A Riemannian manifold $(M,h)$
 \item A smooth vector field $\hat{W} \in \Gamma(TSU(n))$ on $SU(n)$ such that $\pounds_{\hat{W}}(h)=0$ (that is, the Lie derivative of the metric is $0$, or equivalently $\hat{W}$ is a Killing vector field).
\end{itemize}
Given that $F$ is the Randers metric solving the Zermelo navigation problem on $M$ for $h$ and $\hat{W}$, then the unit $F$ speed geodesics of $F$ are given by $\hat{V}_t = \phi_t(\hat{S}_t)$, where:
\begin{itemize}
 \item $\phi_t$ is the flow associated to $\hat{W}$
 \item $\hat{S}$ is a unit speed geodesic of $h$
\end{itemize}
\end{theorem}
Any geodesic of $F$ obtained this way is a length minimiser if and only if the associated Riemannian geodesic of $h$ is a length minimiser of $h$ \cite{CRob}.

In the case that $h$ is the bi-invariant metric, the unit speed geodesics $\hat{S}_t$ are the one parameter subgroups of $SU(n)$, parameterised to have unit $h$ speed.
These can all be expressed as $\hat{S}_t = \exp (i t \hat{D})$ for some $\hat{D} \in \mathfrak{su}(n)$ that is a unit vector for the same $h$.
The flow associated to the vector field $\hat{W}_{\hat{U}} = -i\hat{H}_0 \hat{U}$ is $\phi_t(\hat{U}) = \exp(-it\hat{H}_0)\hat{U}$.
This follows from the observation that the equation defining the flow is exactly the Schr\"odinger equation with Hamiltonian $\hat{H}_0$.
We thus conclude that the time optimal trajectories are given by:
\begin{align}
\label{geos2}
 \hat{U}_t & = \phi_t(\hat{S}_t) = \phi_t\left(\exp (i t \hat{D})\right) \\
	   & = \exp(-it\hat{H}_0) \exp (i t \hat{D} ) \nonumber
\end{align}
This is to be compared with \cite[eqn.51]{ACAR2} which exhibits a similar product of exponential structure.

We determine the optimal Hamiltonian by assuming $\hat{U}_t$ solves the Schr\"odinger equation for an as yet unknown Hamiltonian $\hat{H}_t$:
\begin{align}
 \frac{d \hat{U}_t}{dt} = -i\hat{H}_t \hat{V}_t
\end{align}
which implies that:
\begin{align}
 \hat{H}_t & = i \frac{d \hat{U}_t}{dt} \hat{U}_t^{\dagger} \\
	   & = i \left( (-i\hat{H}_0) \hat{V}_t + \hat{V}_t (i \hat{D}) \right) \hat{V}^{\dagger} \nonumber \\
	   & = i \left( (-i\hat{H}_0) + \hat{V}_t (i\hat{D}) \hat{V}^{\dagger} \right) \nonumber \\
	   & = \hat{H}_0 - \hat{V}_t (\hat{D}) \hat{V}^{\dagger} \nonumber \\
	   & = \hat{H}_0 - \exp(-it\hat{H}_0) (\hat{D}) \exp(it\hat{H}_0) \nonumber \\
	   & = \hat{H}_0 + i \text{Ad}_{\exp(-it\hat{H}_0)}(i \hat{D}) \nonumber
\end{align}

In order to conclude that these are the geodesics and their associated Hamiltonians, we check that the given $\hat{W}_{\hat{U}}$ is a Killing field for the metric $h$.
This is achieved by checking $\pounds_{\hat{W}}(h) = 0$ thus:
\begin{align}
& \frac{d}{dt} h_{\phi_t(\hat{V})}\left(d\phi_t \big|_{\hat{V}}(\hat{A}\hat{V}), d\phi_t \big|_{\hat{V}}(\hat{A}\hat{V})\right) \bigg|_{t=0} \\
&  = \frac{d}{dt} h_{\exp(-it\hat{H}_0)\hat{V}}\left(\exp(-it\hat{H}_0)\hat{A}\hat{V}, \exp(-it\hat{H}_0)\hat{A}\hat{V}\right) \bigg|_{t=0} \nonumber \\
& = \frac{d}{dt} h_{\exp(-it\hat{H}_0)}\left(\exp(-it\hat{H}_0)\hat{A}, \exp(-it\hat{H}_0)\hat{A}\right) \bigg|_{t=0} \nonumber \\
& = \frac{d}{dt} h_{\hat{I}} (\hat{A}, \hat{A} ) \bigg|_{t=0} = 0 \nonumber
\end{align}
where $\hat{V}$ is an arbitrary group element and $\hat{A} \hat{V}$ is an arbitrary element of $T_{\hat{V}}SU(n)$.
We have $d\phi_t \big|_{\hat{V}}(\hat{A}\hat{V}) = \exp(-it\hat{H}_0) \hat{A}\hat{V}$ trivially, as it is the differential of a linear map.
Here, both the left and the right invariance of the metric $h$ have been appealed to;
this proof would need to be modified, or may not hold, in the case that $h$ is not the unique bi-invariant metric.

We conclude (by subtracting $\hat{H}_0$) that the control Hamiltonian $\hat{H}_c(t)$ driving a system (which meets the required premises) along a time optimal trajectory is given by:
\begin{align}
\label{eqn:opham}
\hat{H}_c(t) = - \exp(-it\hat{H}_0) \hat{D} \exp(it\hat{H}_0)
\end{align}
This is constant exactly when $\hat{D}$ commutes with $\hat{H}_0$, which, from eqn.(\ref{nothanks}), is equivalent to saying that $\hat{H}_0$ commutes with $\hat{O}$.
Thus we have obtained a necessary and sufficient condition for the time optimality of constant control fields for any system meeting the premises of the above derivation.
This condition neatly sidesteps the need for the analysis of ``homogenous geodesics'' (see \cite{HOMGEO, Dar2} for much interesting mathematical discussion and \cite{thi, mars} for more direct physical applications of the concept) of right invariant Randers metrics, which could pose significant mathematical challenges.
See section (\ref{HomGeoOCC}) for a fuller discussion of this topic of homogeneous geodesics.

What remains to determine is the formula for a geodesic with desired endpoints (connecting the identity $\hat{I}$ to a desired operator $\hat{O} \in SU(n)$) and the corresponding Hamiltonian.
This boils down to determining the $\hat{D}$ corresponding to a given $\hat{O} \in SU(n)$.
To determine which $\hat{D}$ yields the geodesic with endpoints $\hat{I}$ and $\hat{O}$ such that the system traverses the geodesic in time $T$, we need to solve:
\begin{align}
\label{optraj}
 \hat{U}_T = \exp(-iT\hat{H}_0) \exp (iT \hat{D}) = \hat{O}
\end{align}
Rearranging and taking logs:
\begin{align}
\label{nothanks}
 \exp (iT \hat{D}) = \exp(iT\hat{H}_0) \hat{O} \\
 i\hat{D} = \frac{1}{T}\log\left( \exp(iT\hat{H}_0)\hat{O} \right)
\end{align}
which yields the desired geodesic and corresponding control Hamiltonian:
\begin{align}
\label{eqn:ultans}
 \hat{U}_t & = \exp(-it\hat{H}_0) \exp \left(\frac{t}{T}\log\left( \exp(iT\hat{H}_0)\hat{O} \right) \right) \\ 
	   & = \exp(-it\hat{H}_0) \left( \exp(iT\hat{H}_0)\hat{O} \right)^{{t}/{T}} \nonumber
\end{align}
\begin{align}
 \hat{H}_c(t) & = \frac{i}{T} \exp(-it\hat{H}_0) \log\left( \exp(iT\hat{H}_0)\hat{O} \right) \exp(it\hat{H}_0) \\
	      & = \frac{i}{T} \log\left(\exp(-it\hat{H}_0) \exp(iT\hat{H}_0)\hat{O} \exp(it\hat{H}_0) \right) \nonumber \\
	      & = \frac{i}{T} \log\left(\exp\left(i(T-t)\hat{H}_0\right)\hat{O} \exp(it\hat{H}_0) \right) \nonumber
\end{align}
We can take the $\exp(\pm it\hat{H}_0)$ factors inside the logarithm,
because the matrix logarithm is analytic \cite[Ch.7]{mlog},
which follows from the fact that any matrix function $f$ which is defined by a power series obeys $f(\hat{V}^{-1}\hat{A} \hat{V}) = \hat{V}^{-1} f(\hat{A}) \hat{V}$ for all matrices $\hat{A}$ and all non singular $\hat{V}$.

\subsection{Optimal Times}

Now we attempt to find $T_{\text{opt}}$, the optimal time for implementing some given $\hat{O}$.
Insisting that the left hand side of eqn.(\ref{nothanks}) has norm $1$ according to $h$, that $h(i \hat{D}, i \hat{D}) = 1$, in accordance with the premise that it is the \emph{unit speed} geodesics of $h$ that are needed, we determine that:
\begin{align}
1 & = h(i \hat{D}, i \hat{D}) = h\left(\frac{1}{T}\log\left( \exp(iT\hat{H}_0)\hat{O} \right), \frac{1}{T}\log\left( \exp(iT\hat{H}_0)\hat{O} \right)\right) 
\end{align}
which yields the following equation to be solved for $T$:
\begin{align}
\label{worse}
-\frac{\kappa}{T^2} \Tr \left( \left[ \log\left(\exp(iT\hat{H}_0)\hat{O} \right) \right]^2 \right) = 1
\end{align}
The smallest positive solution that is truly the optimal time; we refer to this as $T_{\text{opt}}$.
At the time of writing, we have found no method for solving this analytically in general; it appears prohibitively difficult by standard means known to the authors.
However, once $\hat{H}_0$ and $\hat{O}$ are given, it can easily be solved numerically; some simple cases are illustrated in \S\ref{exams}.

One special case that can be solved analytically is where $\hat{O}$ and $\hat{H}_0$ commute.
Expanding the matrix logarithm using $\log(\hat{A}\hat{B}) = \log(\hat{A}) + \log(\hat{B})$, rearranging and applying the standard quadratic formula gives:
\begin{align}
 T_{\text{opt}} = \frac{i \kappa \Tr (\hat{H}_0\log (\hat{O}))}{\kappa\Tr(\hat{H}_0^2) -1} 
                   \pm \sqrt{\frac{\kappa\Tr((\log (\hat{O}))^2)}{\kappa\Tr(\hat{H}_0^2) -1}
                            -\frac{\kappa^2(\Tr(\hat{H}_0\log (\hat{O}))^2)}{(\kappa\Tr(\hat{H}_0^2) -1)^2}}
\end{align}
where, as in \cite{mememe}, the $\pm$ is chosen to ensure a positive time.

Once $T_{\text{opt}}$ is known, either analytically or numerically, then the true geodesics and corresponding optimal control Hamiltonian are:
\begin{align}
\label{yeah1}
 \hat{U}_t & = \exp(-it\hat{H}_0) \left( \exp(iT_{\text{opt}}\hat{H}_0)\hat{O} \right)^{{t}/{T_{\text{opt}}}} \\
 \hat{H}_c(t) & = \frac{i}{T_{\text{opt}}} \log\left(\exp\left(i(T_{\text{opt}}-t)\hat{H}_0\right)\hat{O} \exp(it\hat{H}_0) \right) \nonumber
\end{align}

We can use the well-known BCH formula \cite[\S3]{hall} to evaluate approximations to $i\hat{D}$ as it provides a series type representation for the solution to
 $\exp(z) = \exp(x)\exp(y)$.
Given a certain $x,y$ in a Lie algebra the solution for $z$ is given by:
\begin{align}
z = x + y + \frac{1}{2}[x,y] + \frac{1}{12} [x,[x,y]] - \frac{1}{12}[y,[x,y]] - \frac{1}{24}[y,[x,[x,y]]] + \cdots
\end{align}
We apply this to eqn.(\ref{nothanks}) to solve
$\exp(iT \hat{D}) = \exp(iT\hat{H}_0) \exp(\log(\hat{O}))$,
to obtain:
\begin{align}
i \hat{D} &  = i\hat{H}_0 + \frac{1}{T}\log{\hat{O}} + \frac{1}{2}[i\hat{H}_0, \log(\hat{O})] + \frac{T}{12}[i\hat{H}_0, [i\hat{H}_0, \log(\hat{O})]] \\
& \qquad - \frac{1}{12}[\log{\hat{O}}, [i\hat{H}_0, \log{\hat{O}}]] - \frac{T}{24}[\log(\hat{O}), [i\hat{H}_0, [i\hat{H}_0, \log(\hat{O})]]] + \cdots \nonumber
\end{align}

\subsection{Scope of Our Approach}

The choice that $h$ is the bi-invariant metric in the above derivation of the geodesics given in eqn.(\ref{geos}) allows us to exploit the fact that the geodesics of this metric are the one parameter subgroups.
If a different right invariant metric $h_{\hat{U}}$ were chosen (that is, a different physical constraint on $\hat{H}_c(t)$), then the geodesics of $h_{\hat{U}}$ would need to be found by a different method.
In the case that, at the identity, $h$ (in some basis of $\mathfrak{su}(n)$) has the form $h_{ij}$, then its unit speed geodesics can be found using the standard methods of the Riemannian geodesic equation (or the EP equations for a right invariant Riemannian metric), although they will not necessarily be the one parameter subgroup.

This section depends on being able to apply theorem (\ref{great}) to determine the geodesics of $F$ in terms of the geodesics of $h$ and the flow $\phi_t$ associated to $\hat{W}_{\hat{U}} = -i\hat{H}_0\hat{U}$.
In order to do this we need that $\hat{W}_{\hat{U}}$ is a Killing vector field of $h$.
In such cases the following results apply.
At the time of writing, we are unaware of tractable, necessary and sufficient conditions for checking when a right invariant Riemannian metric on $SU(n)$ has any right invariant Killing fields; it is well known that all \emph{left} invariant vector fields are Killing for such a metric.
However, there are certainly at least some in addition to the bi-invariant metric, for example:
\begin{align}
 g_{\hat{U}}(\hat{A}\hat{U}, \hat{A}\hat{U}) = \Tr(\hat{A}^{\dagger}\hat{A}) + \zeta \Tr(\hat{A}, \hat{C})^2
\end{align}
for $\zeta \in \mathbb{R}^{+}$ small enough that the formula for $g$ defines a positive definite Riemannian metric.
It is readily checked that this metric has the right invariant vector field $\hat{X}_{\hat{U}} = \hat{C}\hat{U}$ for a Killing field.

The unit speed geodesics of an arbitrary $h$ can be found by using the Euler Poincar\'e equation:
\begin{align}
 \frac{d}{dt} \frac{ \left( \partial h_{ij}\xi^i\xi^j \right)}{\partial \xi^d} = -C^{a}_{b d}\frac{\left(\partial h_{ij} \xi^i \xi^j \right)}{\partial \xi^a} \xi^b
\end{align}
This yields:
\begin{align}
 \dot{\xi}^k = -C^{a}_{bd} h^{kd} h_{a i} \xi^i \xi^b
\end{align}

When this can be solved, one can obtain the unit speed geodesics of $h$ and apply theorem (\ref{great}) again.
Once the geodesics of $h$ are known they can be used to find the geodesics of the Randers metric $F$ solving a Zermelo navigation problem on $SU(n)$ where the `wind' is the vector field $\hat{W}_{\hat{U}}=-i\hat{H}_0 \hat{U}$.
If the unit speed geodesics of $h$ are given by $\hat{S}_t$, then the geodesics of $F$ are given by $\hat{V}_t = \phi_t(\hat{S}_t) = \exp(-it\hat{H}_0) \hat{S}_t$.

We can now solve for the optimal Hamiltonian in a manner similar to the specific case where $h$ is the bi-invariant metric, by setting:
\begin{align}
 \frac{d \hat{V}_t}{dt}=-i\hat{H}_t \hat{V}_t
\end{align}
and deducing that the following is required:
\begin{align}
 \frac{d}{dt} \left( \exp(-it\hat{H}_0)\hat{S}_t \right) = -i\hat{H}_t \exp(-it\hat{H}_0)\hat{S}_t
\end{align}
which yields:
\begin{align}
\label{optimalham}
\hat{H}_t = \hat{H}_0 - \exp(-it\hat{H}_0)\left(-i\frac{d\hat{S}_t}{dt}\hat{S}_t^{\dagger} \right) \exp(it\hat{H}_0)
\end{align}
If $\hat{S}_t$ are the unit speed geodesics of $h$, then by solving the EP equation associated to $h$, we can find the Hamiltonian driving along $\hat{S}_t$ at unit speed.
Suppose then that this Hamiltonian is known and solves:
\begin{align}
 \frac{d\hat{S}_t}{dt} = -i\hat{Q}_t\hat{S}_t
\end{align}
It follows that:
\begin{align}
 \hat{Q}_t = i\frac{d\hat{S}_t}{dt}\hat{S}_t^{\dagger}
\end{align}
Substituting into eqn.(\ref{optimalham}) gives:
\begin{align}
\hat{H}_t & = \hat{H}_0 - \exp(-it\hat{H}_0)(i\hat{Q}_t ) \exp(it\hat{H}_0)
\end{align}
The control Hamiltonian is:
\begin{align}
\hat{H}_c(t)  = \exp(-it\hat{H}_0)(i\hat{Q}_t ) \exp(it\hat{H}_0) 
\end{align}
All that is required to obtain the optimal trajectories is to solve the EP equation for a right invariant Riemannian metric on $SU(n)$ and then apply the above procedure.
This can be achieved by solving a first order system of ODEs.
As this system is the EP equation for a \emph{quadratic} $\ell$, it is possible to solve for the time derivative of $\xi$ analytically.
This makes numerical integration even simpler.

Solving for a geodesic with specified end points appears difficult without first actually obtaining the $h$ geodesics in closed form.
However, in some simplifying cases of the value of $h$ the unit speed geodesics can be obtained in closed form using Jacobi Elliptic functions.
For an example of how to do this on the group $SO(3)$,
see any mathematical mechanics textbook (eg \cite{mars}) that includes a derivation of the solutions of the Euler equations for a falling rigid body.
The $SU(2)$ case proceeds similarly and is tractable, essentially because $\mathfrak{so}(3) \cong \mathfrak{su}(2)$, as is familiar from the theory of spin half particles under spatial rotations in quantum mechanics.

\subsection{Physical Constraints Encompassed and not Encompassed by Shen's Theorem}
\label{prob}

It is worth noting what is {\it not} achieved by this approach.
The only constraints that can be studied this way are those which restrict the control Hamiltonian $i\hat{H}_c$ to the unit sphere of a norm induced by a given inner product $h$ on $\mathfrak{su}(n)$.
This is because such inner-products are in one-to-one correspondence with right invariant metrics on $SU(n)$ (by right translation).
This allows only quadratic constraints to be studied.
An interesting constraint not included in this class is the restriction that the energy expectation (in some specific state $| \psi \ra$) associated to the control Hamiltonian alone is equal to a fixed constant for all time:
\begin{align}
 \kappa \la \psi | \hat{H}_{c}(t) - \frac{1}{n} E_0(t)\hat{I}| \psi \ra = 1
\end{align}

Future work will include potentially applying the results in \cite{FinNav, Shen} to a generalisation of the setup described here.
The desired generalisation would be to relax the condition that the function representing the constraint on the control Hamiltonian is an inner product, and to allow more general Minkowski norms to take this role instead.
This leads to a desire to solve the problem of Zermelo navigation on Finsler manifolds rather than Riemannian ones, which is currently not solved.
See \cite{Bao2004} for an exhaustive account of the status of Zermelo navigation of Riemannian manifolds.

The main result of \cite{FinNav} is particularly relevant as it generalises \cite{CRob} and allows one to replace the role of $h$ with a general Finsler metric that has $i\hat{H}_0\hat{U}$ as a Killing field.
One class of Finsler metrics with this property is the bi-invariant ones, of which there are many.
The proof is similar to the bi-invariant Riemannian cases already presented, and so is omitted.
Examples of such constraints are found in the Finsler metrics formed from the right translation of the Shatten $p$-Norms on $\mathfrak{su}(n)$.
These correspond the constraint that
$\label{spn} F^{(p)}(\hat{H}_c(t)) := \kappa\left(\sum_n |E_n|^p\right)^{1/p} =  1$ $\forall t$, thus generalising the case of the bi-invariant Riemannian metric studied above, wherein $p=2$ (the only value of $p$ yielding a Riemannian metric on $SU(n)$).
Solving the navigation problem in general has not been achieved by the mathematics community, as far as the authors are aware.
However, there are other cases besides the Riemannian case that have been solved; the Kropina metric case \cite{Yosh} is notable.
In the absence of a solution to the navigation problem analogous to the role of Randers metrics in the Riemannian case, alternative methods must been sought.
The central result of \cite{FinNav} allows one to determin the geodesics of the Finsler metric solving the navigation problem on $SU(n)$ for which $\hat{H}_c(t)$ is constrained such that $F^{(p)}(\hat{H}_c(t))=1$.
The geodesics of the (yet unknown) metric solving the navigation problem in such a generalised case are:
\begin{align}
 \hat{V}_t = \exp(-it\hat{H}_0) \exp (it \hat{D})
\end{align}
where $F^{(p)}(i\hat{D})=1$.

One can also find the time optimal Hamiltonian for implementing gate $\hat{O}$ to be the same as the Riemannian ($p=2$) case:
\begin{align}
 \hat{H}_c(t) = \frac{i}{T} \exp(-it\hat{H}_0) \log\left( \exp(iT\hat{H}_0)\hat{O} \right) \exp(it\hat{H}_0)
\end{align}
except with $T_{\text{opt}}$ taking a different value.
The requirement is now that $T_{\text{opt}}$ is the value of $T$ that solves:
\begin{align}
 F^{(p)}\left(\log(\exp(iT\hat{H}_0)\hat{O})\right) = T
\end{align}

Further work will also include an investigation into application of \cite{NewFins}.
This work extends the results of \cite{FinNav, CRob} to so called 'conic'-Finsler metrics and obtains a similar form for the geodesics under broadly analogous conditions.

\subsection{Homogeneous Geodesics and the Optimality of Constant Controls}
\label{HomGeoOCC}

A \emph{homogeneous geodesic} (through the identity) on a Lie group with a Finsler metric $F$ is a one-parameter subgroup which is also a geodesic \cite{HOMEGEO2,HOMGEO}.
Here, it is a curve of the form $\hat{U}_t = \exp(-it\hat{A})$ for some $i\hat{A} \in \mathfrak{su}(n)$, which is also a geodesic of $F$.
The $\hat{A} \in \mathfrak{su}(n)$ is a \emph{geodesic vector}.
These are exactly the curves that can be trajectories of a controlled quantum systems (of the type discussed throughout) for which only constant controls are permitted, as discussed in \cite{mememe}.

Theorem 3.1 in \cite{HOMGEO} presents a condition, which need to be mildly adapted to our situation.
Adapted to the present example of $SU(n)$ (rather than a more general homogeneous space), the condition for $\hat{X}$ to be a geodesic vector is:
\begin{align}
\label{thmHG}
 g_{\hat{X}} (\hat{X} , [\hat{X}, \hat{Z}] ) = 0, \forall \hat{Z} \in \mathfrak{su}(n)
\end{align}
where $g$ is the fundamental tensor of $F$ restricted to $\mathfrak{su}(n)$.
Here the fundamental tensor of a Finsler metric is the Hessian of the same metric point-wise.
For the definition of the fundamental tensor, much more detailed discussion of its role in Finsler geometry, and many specifics of Randers metrics, see \cite{ran}.

Many of the results about homogeneous geodesics of Finsler metrics on Lie groups are applicable to \emph{left} invariant metrics.
However, these results can be easily adapted to the right invariant case, which arises naturally in quantum mechanics.
The construction of an ``opposite group'' allows one to adapt results without difficulty; typically only sign changes are incurred.

Theorem 3.7 in \cite{HOMGEO} can be applied to establish that any right invariant Randers metric on $SU(n)$ ($n\geq2$) will have infinitely many homogeneous geodesics.
The theorem establishes that a right invariant Finsler metric on a compact Lie group will have infinitely many geodesic vectors.
This can be easily seen by direct application of the theorem and by the observation that the rank of $SU(n)$ is $n - 1$, and thus the theorem applies to any qubit system of more than one bit.
The theorem requires that the rank is $\geq 2$, thus in all cases except $SU(2)$ (the single qubit), there exist infinitely many homogeneous geodesics.
This establishes the importance of determining all homogeneous geodesics of right invariant Randers metrics on $SU(n)$.
Furthermore, in the case of a Randers metrics (as is our case) theorem (4.2) in \cite{HOMGEO} establishes a practically simplifying condition on homogeneous geodesics.

\section{Example Calculations}

\label{exams}

In this section we use the results derived above to calculate speed limits; these follow the examples in \cite{mememe}, except here allowing time dependent controls.
Throughout this section the metric $h$ representing the constraint on $\hat{H}_c(t)$ is $h(i\hat{A},i\hat{B})= \kappa \Tr((i\hat{A})^{\dagger} i\hat{B}) 
= \kappa \Tr(\hat{A}^{\dagger}\hat{B})
= \kappa \Tr(\hat{A}\hat{B})$.
The dagger vanishes in the final step since $\hat{A}$ is Hermitian.
This yields the constraint $\Tr(\hat{H}_c(t)^2) = \frac{1}{\kappa}$ for all times.

\subsection{Single Spin in a Magnetic Field}
\subsubsection{Pauli $\sigma_y$ Gate}

Following the example calculations performed in \cite{mememe} we study the speed limit to performing the gate:
\begin{align}
\hat{O} =
\begin{pmatrix}
 0 & -1 \\
 1 & 0 
\end{pmatrix}
= -i\sigma_y
\end{align}
in a system constrained such that $h\left((i\hat{H}_c(t))^{\dagger} i\hat{H}_c(t)\right) = \kappa \Tr(\hat{H}_c(t)^2) = 1$ with drift Hamiltonian:
\begin{align}
 \hat{H}_0 = B^{x} \sigma_x + B^{y} \sigma_y
\end{align}
such that $h(i\hat{H}_0, i\hat{H}_0) \leq 1$.

Substituting these conditions into eqn.(\ref{worse}) for the optimal time gives
\begin{align}
\label{bad}
-\frac{\kappa}{T^2} \Tr \left(\log 
(\mu(T))^2 
\right) = 1
\end{align}
where the matrix $\mu$ is defined as
\begin{align}
\mu(T) :=  
\begin{pmatrix}
\frac{\displaystyle\sin(TB)(B^y+iB^x)}{\displaystyle B}
    & -\cos(T B) \\ 
\cos(T B) 
    & \frac{\displaystyle\sin(T B)(B^y-iB^x)}{\displaystyle B} 
    \end{pmatrix}
\end{align}
and where $B = ||\vec{B}||$ (standard Euclidean norm). 

A numerical solution can be achieved using any zero finding algorithm; we found Mathematica's ``FindRoot'' method to be  effective when given suitable starting points.
The result (due to the cyclic property of the matrix trace) depends on the eigenvalues of the quantity within the trace.
Thus one should diagonalize before attempting to numerically solve eqn.(\ref{worse}).
It greatly simplifies the process and can be easily achieved with any good algebra package.

By eqn.(\ref{eqn:ultans}), the geodesic (of the Randers metric on $SU(2)$ which solves Zermelo's navigation problem) which connects $\hat{I}$ to $\hat{O}$ is given by:
\begin{align}
 \hat{V}_t & = \exp(-it\hat{H}_0) \exp \left(\frac{t}{T_{\text{opt}}}\log\left( \exp(iT_{\text{opt}}\hat{H}_0)\hat{O} \right) \right) \\ 
           & = \exp(-it\hat{H}_0) \left( \exp(iT_{\text{opt}}\hat{H}_0)\hat{O} \right)^{t/T_{\text{opt}}} \nonumber \\
           & = \exp \left(-it (B^x\sigma_x + B^y\sigma_y ) \right) 
        \left(\vphantom{\strut} \exp(iT_{\text{opt}} (B^x\sigma_x + B^y\sigma_y ))(-i\sigma_y) \right)^{t/T_{\text{opt}}} \nonumber \\
           & = \exp \left(-it \left(B^x\sigma_x + B^y\sigma_y \right) \right) 
        (\mu({T_\text{opt}}))^{t/T_{\text{opt}}} \nonumber
\end{align}
These geodesics are not generally of the form of any time independent trajectory, i.e. the optimal controls are not constant.

The optimal control Hamiltonian in closed form is:
\begin{align}
\hat{H}_c(t) & = \frac{i}{T_{\text{opt}}} \exp(-it \left(B^x\sigma_x + B^y\sigma_y \right)) 
    \log ( \mu({T_\text{opt}}))
    \exp(it \left(B^x\sigma_x + B^y\sigma_y \right)) \nonumber
\end{align}
where
\def\expHO{{\cal E}}
\begin{align}
 \exp(-it \left(B^x\sigma_x + B^y\sigma_y \right)) = 
\begin{pmatrix} \cos(t B) 
    & \frac{\displaystyle-\sin(t B)(B^y + iB^x)}{\displaystyle B} \\ 
    \frac{\displaystyle\sin(t B)(B^y - iB^x)}{\displaystyle B} 
    & \cos(t B) 
    \end{pmatrix}
\end{align}
Expanding these quantities in an attempt to find $\hat{H}_c(t)$ does not lead to tractable expressions.

The optimal control fields can also be determined.
Consider that the control Hamiltonian can be expressed as $i\hat{H}_c(t) = D^{k}_t i\sigma_k$.
One can now extract the individual control fields $D^k$ from the following, as $i\sigma_k$ are an orthogonal basis (according to $h(i\hat{A}, i\hat{B}) = \kappa \Tr(\hat{A}\hat{B}))$ for $\mathfrak{su}(n)$:
\begin{align}
\label{hmm}
 D^k_t = h \left(i\hat{H}_c(t), \frac{i\sigma_k}{2}\right) = \frac{\kappa}{2} \Tr(\hat{H}_c(t) \sigma_k)
\end{align}
This calculation rapidly becomes intractable analytically, however by substituting the Taylor series for $log$ and $exp$, one can find a simple series representation for the optimal control fields.
Numerically extracting the control fields to arbitrary precision is straightforward.

\subsubsection{Specific Physical Instance}

We now specialise to specific values for $\kappa$, $B^x$ and $B^y$ in order to numerically obtain the optimal time.
We set $B^x = B^y = 1/4$ and $\kappa=1$; one readily checks that this system meets the ``small wind'' premise of Shen's theorem.

We solved the specific instance of eqn.(\ref{worse}) using Mathematica.
The smallest, real, positive root is $T_{\text{opt}} \approx 3.2043 \ldots$.
(The actual physical time in seconds can be obtained by reintroducing the physical constants that have been lost after non-dimensionalisation throughout.
Specifically, $\hbar$ has been set to $1$ throughout.)

In this instance the optimal control Hamiltonian is given by
\begin{align}
 \hat{H}_c(t) & = \frac{i}{T_{\text{opt}}} \exp\left(\frac{-it}{4}(\sigma_x + \sigma_y) \right) \log
 \begin{pmatrix} \frac{1+i}{\sqrt{2}} \sin \left(\frac{T_{\text{opt}}}{\sqrt{8}}\right) & -\cos\left(\frac{T_\text{opt}}{\sqrt{8}}\right) \\ \cos\left(\frac{T_\text{opt}}{\sqrt{8}}\right) & \frac{1-i}{\sqrt{2}} \sin \left(\frac{T_{\text{opt}}}{\sqrt{8}}\right) \end{pmatrix}
 \exp\left(\frac{it}{4}(\sigma_x + \sigma_y) \right)
\end{align}
by eqn.(\ref{eqn:ultans}).
We have evaluated the logarithm exactly in closed form, but the result is cumbersome and does not provide any physical insight, so it is omitted.

\begin{figure}[tp]
\centering
\includegraphics[scale=0.3]{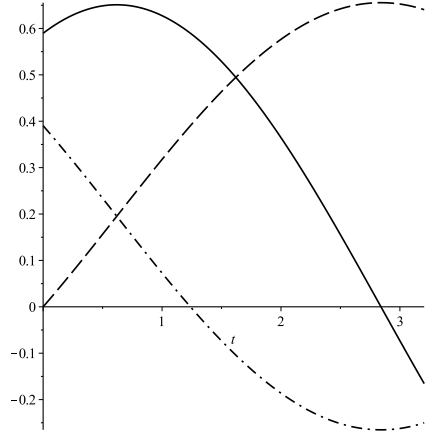}
\caption{Optimal control fields $D^x$ (dashed), $D^y$ (dot-dashed), $D^z$ (solid) for Pauli $y$ gate, as a function of time, in the $B= 1/4$ case. \label{fig:win}}
\end{figure}

We obtain the optimal control fields from this result.
The $k$th field is obtained by evaluating eqn.(\ref{hmm}) $\frac{1}{2}\Tr(\hat{H}_c(t) \sigma_k)$ numerically, as shown in fig.(\ref{fig:win}).
As a check, we have numerically confirmed that the control fields have the property that the sum of their squares is $1/2$ ($\forall t$), which the constraint on $\Tr(\hat{H}_c(t)^2)$ mandates.

\subsubsection{Example of Assessing Quantum Systems}

Given a choice of quantum systems for the potential implementation of fast quantum computation, one requires methods to assess systems.
Here we show how the methods illustrated in this paper can be used to perform such assessment through an example.

As shown in \cite{mememe}, the optimal time-independent control implementation time for the same gate as above $-i\sigma_y$ is:
\begin{align}
 & T_{\text{opt}} = \frac{\pi}{2} \frac{B^y}{(D^2 - B^2)}\left(1 \pm \sqrt{1 + \frac{D^2 - B^2}{(B^y)^2}} \right)
\end{align}
where $D = ||\vec{D}||$ (standard Euclidean norm). 

\begin{figure}[tp]
\centering
\includegraphics[scale=0.5]{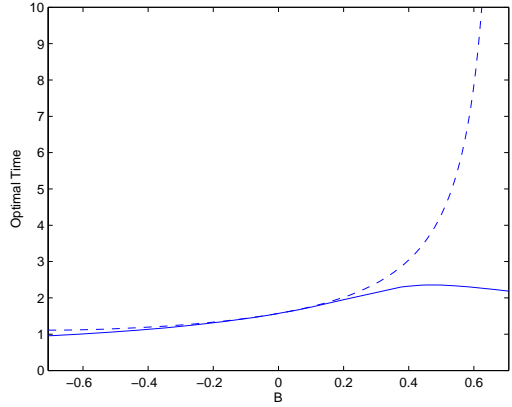}
\caption{Optimal times for time dependent (solid) and independent (dashed) controls. \label{fig:times}}
\end{figure}

To illustrate, we set $\kappa=1/2$ (achieved through setting the value of $D$ = 1, a choice made for ease) and $B^x = B^y = b$ (some real number $b^2 < 1/2$).
Then the optimal time in the time independent case is:
\begin{align}
T_{\text{opt}} = \frac{\pi}{2} \frac{b}{(1 - 2b^2)}\left(1 \pm \frac{\sqrt{1-b^2}}{b} \right)
\end{align}
where the $\pm$ is chosen to make the time always positive \cite{mememe}.
We numerically solve eqn.(\ref{bad}) in Matlab, with the given parameters substituted to find the optimal time (as a function of $b$ also) when time dependent controls are allowed.
This yields the results of fig (\ref{fig:times}).
This illustrates that some systems and some types of controls are significantly better for the implementation of specific QIP tasks.
The method presented is a powerful tool for assessing such situations.
One can clearly see from fig.(\ref{fig:times}) that scenarios with values of $b < 0.2$ are favorable over the $b > 0.2$ region if only constant controls are permitted.

The computation time to obtain the fig.(\ref{fig:times}) and the optimal control schemes shown in fig.(\ref{fig:win}) were both negligible in Matlab, and that this also appears to be the case for two qubit gates.

\subsection{Swap Gate in a Length Two Heisenberg Spin Chain}
\subsubsection{Two Spin Chain}

The drift Hamiltonian for a two spin chain with (arbitrary spin coupling) is \cite{xyzchain}:
\begin{align}
\label{eqn:driftsc}
\hat{H}_0 = J^x \sigma_x \otimes \sigma_x + J^y \sigma_y \otimes \sigma_y + J^z \sigma_z \otimes \sigma_z
\end{align}
Again, we take $h$ to be the Killing form so we can apply theorem (\ref{great}) to obtain the geodesics in closed form.
By eqn (\ref{eqn:ultans}), the geodesic which connects $\hat{I}$ to $\hat{O}$ of the relevant Randers metric on $SU(4)$ are given by eqn. (\ref{yeah1}).
One can exploit the block diagonal form of $\hat{H}_0$ in order to simplify the equation that needs to be solved.

The optimal Hamiltonian is:
\begin{align}
 \hat{H}_c(t) = & \frac{i}{T_{\text{opt}}} \exp\left(-it\left(J^x \sigma_x \otimes \sigma_x + J^y \sigma_y \otimes \sigma_y + J^z \sigma_z \otimes \sigma_z \right)\right)  \\
    & \qquad \times \log\left( \exp\left(iT_{\text{opt}} \left(J^x \sigma_x \otimes \sigma_x + J^y \sigma_y \otimes \sigma_y + J^z \sigma_z \otimes \sigma_z \right)\right)\hat{O} \right) \nonumber \\
 & \qquad \times\exp(it\left(\left(J^x \sigma_x \otimes \sigma_x + J^y \sigma_y \otimes \sigma_y + J^z \sigma_z \otimes \sigma_z \right)\right)) \nonumber
\end{align}
The actual control fields can be extracted, similarly to before, via:
\begin{align}
\label{cfs}
 f_{mn}(t) = \frac{\kappa}{4}\Tr\left(\hat{H}_c(t) \sigma_m \otimes \sigma_n \right)
\end{align}
This is the expansion of $\hat{H}_c(t)$ in a basis for $\mathfrak{su}(4)$.
This basis is:
\begin{align}
\label{bas}
 \{i\sigma_n \otimes \sigma_m \| n,m=0,x,y,z \text{ but not both } n=0 \text{ and } m=0  \}
\end{align}
Here $\sigma^0$ is taken to be the $2\times2$ identity matrix whereas the other $\sigma$s are all the standard Pauli matrices.
One readily checks that this basis is orthogonal w.r.t. the Killing form, which is the key property applied when extracting the control fields in eqn.(\ref{cfs}).
The origin of the $4$ in this formula is the trace of the $4\times4$ identity matrix.
Explicitly, we are representing $\hat{H}_c(t)$ as:
\begin{align}
 \hat{H}_c(t) = \sum f_{mn}(t)\sigma_m\otimes \sigma_n
\end{align}
where ths sum is over the basis vectors appearing in eqn (\ref{bas}).

\subsubsection{Specific Physical Instance: The $XXX$-Spin Chain}

\begin{figure}[tp]
\centering
\includegraphics[scale=0.5]{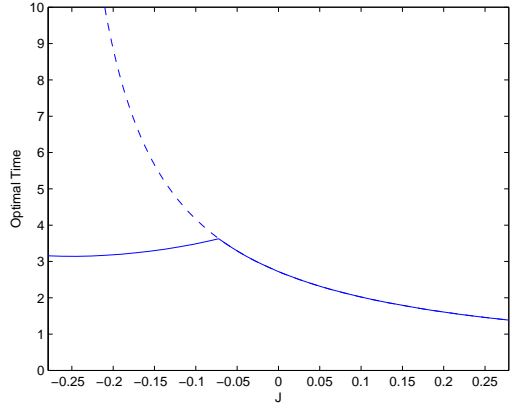}
\caption{Optimal times for time dependent (solid) and time independent (dashed) controls in an XXX Spin Chain. \label{fig:times2}}
\end{figure}

In order to again illustrate the way in which out method allows us to determine which systems are best suited to quickly implementing a QIP task, we study the case of the isotropic Heisenberg spin chain,
 the $J^x = J^y = J^z = J$ case of eqn.(\ref{eqn:driftsc}).
This leaves only one parameter $J$ to consider, yielding a simple pedagogic example for the method.
We set $\kappa = 1$ and consider the optimal time for implementing the (special unitary) swap gate:
\begin{align}
\hat{O} = e^{i \pi/4}
\begin{pmatrix}
1 & 0 & 0 & 0 \\
0 & 0 & 1 & 0 \\
0 & 1 & 0 & 0 \\
0 & 0 & 0 & 1 
\end{pmatrix}
\end{align}
Using eqn (\ref{eqn:ultans}), the optimal Hamiltonian is:
\begin{align}
\hat{H}_{c}(t) & =  \frac{i}{4T_{\text{opt}}}
 \begin{pmatrix}
  2e^{-itJ} & 0 & 0 & 0 \\
  0 & e^{-itJ} + e^{3itJ} & e^{-itJ} - e^{3itJ} & 0 \\
  0 & e^{-itJ} - e^{3itJ} & e^{-itJ} + e^{3itJ} & 0 \\
  0 & 0 & 0 & 2e^{-itJ} \\
 \end{pmatrix}  \\
 & \qquad \times \log\left( \frac{1}{2} e^{\frac{i \pi}{4}}
 \begin{pmatrix}
  2e^{iT_{\text{opt}}J} & 0 & 0 & 0 \\
  0 & e^{iJT_{\text{opt}}} - e^{-3iJT_{\text{opt}}} & e^{iJT_{\text{opt}}} + e^{-3iJT_{\text{opt}}} & 0 \\
  0 & e^{iJT_{\text{opt}}} + e^{-3iJT_{\text{opt}}} & e^{iJT_{\text{opt}}} - e^{-3iJT_{\text{opt}}} & 0 \\
  0 & 0 & 0 & 2e^{iJT_{\text{opt}}} \nonumber \\
 \end{pmatrix}
 \right) \\
 & \qquad \times \begin{pmatrix}
  2e^{itJ} & 0 & 0 & 0 \\
  0 & e^{itJ} + e^{-3iJt} & e^{itJ} - e^{-3itJ} & 0 \\
  0 & e^{itJ} - e^{-3iJt} & e^{itJ} + e^{-3itJ} & 0 \\
  0 & 0 & 0 & 2e^{itJ} \\
 \end{pmatrix} \nonumber 
\end{align}
We can determine the optimal time $T_{\text{opt}}$ as before, by numerically solving eqn (\ref{worse}).
Substituting the specifics of the current problem into this equation yields:
\begin{align}
-\frac{1}{(T_{\text{opt}})^2} \Tr \left( \log\left( \frac{1}{2} e^{\frac{i \pi}{4}}
 \begin{pmatrix}
  2e^{iT_{\text{opt}}J} & 0 & 0 & 0 \\
  0 & e^{iJT_{\text{opt}}} - e^{-3iJT_{\text{opt}}} & e^{iJT_{\text{opt}}} + e^{-3iJT_{\text{opt}}} & 0 \\
  0 & e^{iJT_{\text{opt}}} + e^{-3iJT_{\text{opt}}} & e^{iJT_{\text{opt}}} - e^{-3iJT_{\text{opt}}} & 0 \\
  0 & 0 & 0 & 2e^{iJT_{\text{opt}}} \nonumber \\
 \end{pmatrix}
 \right)^2 \right) = 1
\end{align}
Using the same procedure as in the previous example, we obtain the optimal execution times.
These are shown in fig.(\ref{fig:times2}).
The dashed line in fig.(\ref{fig:times2}) plots the function, adapted to this specific scenario, from \cite[eqn.21]{mememe}, giving the time independent optimal time:
\begin{align}
T_{opt} = \frac{\pi}{2} \left( \frac{\sqrt{3}}{2\sqrt{3}J \pm 1} \right)
\end{align}
where the $\pm$ is again chosen to make the time positive.
The cusp in the time dependent line is due to its behaviour intersecting 
the time independent line, and then following the time independent behaviour.

\subsubsection{Example of Assessing Quantum Systems}

It can be seen from fig.(\ref{fig:times2}) that, in the region between $J\approx-0.075$ and $J=\frac{1}{\sqrt{12}}$, constant controls perform exactly as well as time dependent ones.
However, outside this region, time dependent controls significantly out-perform constant ones for which the optimal time has an asymptote at $J=-\frac{1}{\sqrt{12}}$.
This indicates that our method can be used to assess the choice of the best type of control scheme as well as to determine optimal control fields when a specific physical system is considered.

The fact that the optimal times agree for both types of control scheme on a contiguous region of values for the parameter $J$ suggests that the theorem leading to eqn.(\ref{thmHG}) could potentially be strengthened (perhaps only in the Randers case) to show that the infinitely many geodesics vectors of a Finsler metric are a connected set.
We conjecture that this is the case for Randers metrics on $SU(n)$ ($n \geq 2$).
These results justify the physical relevance of that theorem.

\section{Further Constraints on $\hat{H}_c$ Using Lagrange Multipliers}

\subsection{The Need for Further Constraints}

The constraints studied thus far are, alone, insufficient for many physical applications.
The assumption thus far as been only that the control Hamiltonian is constrained to be such that $h(i\hat{H}_c(t), i\hat{H}_c(t)) = 1$.
In comparison to \cite{ACAR2}, only the roles of the $L_T$ and $L_S$ parts of the Lagrangian have been treated here.
The $L_T$ part is analogous to our application of the result of \cite{Shen} on Zermelo navigation in the case of a right invariant Riemannian metric.
The $L_S$ part has no analogue as our work expresses the problem in a geometrically intrinsic way.
We consider this to be an advantage of our method as it allows a more mathematical view of the problem; intrinsic geometry has been proven many times to be superior for mathematical analysis of geometric problems over methods based on many constraints or specific coordinate systems.
This allows us to formulate the problem as a \emph{first order} system of ODE from the outset by using the EP equation, rather than needing to compute any first variations or use the EL equations.
So we can obtain a first order equation for the optimal Hamiltonian, \emph{without} the need to actually determine any geodesics a priori.

The disadvantage of our method compared with \cite{ACAR2} is that we can handle fewer constraints, as described above (\S\ref{prob}).
Our method thus far can handle only the cases where the ``size'' type constraint \cite{ACAR2} is representable by an inner product.

To motivate the need for further constraints, we again consider the drift Hamiltonian for a two spin ``chain'' (with anisotropic couplings $J$) \cite{xyzchain}:
\begin{align}
\hat{H}_0 = J_x \sigma^x \otimes \sigma^x + J_y \sigma^y \otimes \sigma^y + J_z \sigma^z \otimes \sigma^z
\end{align}
Simply constraining the control Hamiltonian to be such that
 $h(i\hat{H}_c(t),i\hat{H}_c(t)) = 1$
for some inner product $h$ is insufficient for practical applications where the producible set of control Hamiltonians does not include every direction within $\mathfrak{su}(n)$. 
For example, a common model of a controlled spin chain is one in which the control Hamiltonian takes the form:
\begin{align}
 \hat{H}_c(t) = f_1(t) \sigma_z \otimes \hat{I} + f_2(t) \hat{I} \otimes \sigma_z
\end{align}
That is, there is one local control field in the $z$ direction only for each site in the chain.
In such a situation no terms like $\sigma^x \otimes \sigma^x$ (or multiples there of) could appear in the control Hamiltonian, as these represent the couplings between sites in the chain, and are not the effect of \emph{any possible} external field.
In this case (choosing $h$ to be $\kappa$ times the Killing form), the constraint $h(i\hat{H}_c(t), i\hat{H}_c(t)) = 1$ evaluates to $\kappa \Tr(\hat{H}_c(t)^2) = 1$, which only constrains the sum of the squares of the control fields.
An extra constraint must be added to exclude those terms from the control Hamiltonian that cannot be physically implemented.

This can be done by including Lagrange multipliers to create a new functional:
\begin{align}
 \Lambda \left(\hat{U}_t, \frac{d \hat{U}_t}{dt}, \omega_k(t) \right) = \frac{1}{2} \left[F_{\hat{U}_t} \left(\frac{d \hat{U}_t}{dt} \right) \right]^2 + \sum_k \omega_{k}(t) \left(f_{k, \hat{U}_t} \left( \frac{d \hat{U}_t}{dt} \right) - c_k\right)
\end{align}
where $F$ is the Randers metric solving the relevant navigation problem,
$f_k$ represent the additional constraints, and $\omega_k$ are the Lagrange multipliers.
The values $c_k$ represent the value of $f$ to which the trajectory is constrained.

We consider only the case where $f$ is right invariant; this results in $\Lambda$ also being right invariant.
This corresponds to situations where the additional constraints depend only on the Hamiltonian, rather than the on location of $\hat{U}_t$ on $SU(n)$.

In this situation $\Lambda$ can be expressed as:
\begin{align}
\label{eqnfin}
 \Lambda \left(\frac{d \hat{U}_t}{dt} \hat{U}_t^{\dagger}, \omega_k(t) \right) & = \frac{1}{2} \left[F\left(\frac{d \hat{U}_t}{dt} \hat{U}_t^{\dagger} \right) \right]^2 + \sum_k \omega_{k}(t) \left(f_{k} \left( \frac{d \hat{U}_t}{dt} \hat{U}_t^{\dagger} \right) - c_k\right) \\
 & = \frac{1}{2} \left[F\left(-i\hat{H}(t) \right)\right]^2 + \sum_k \omega_{k}(t) \left(f_{k} \left( -i\hat{H}(t) \right) - c_k\right)
\end{align}
where $\hat{H}_t$ is the Hamiltonian such that $\hat{U}_t$ solves the corresponding Schr\"odinger equation.

\subsection{Forbidden Directions}

One specific set of $f_k$ and $c_k$ with practical relevance is:
$f_{k}\left(\frac{d \hat{U}_t}{dt} \hat{U}_t^{\dagger}\right) = \Tr\left(\left((\frac{d \hat{U}_t}{dt} \hat{U}_t^{\dagger} + i\hat{H}_0)i\hat{F}_k \right)\right)$ with $c_k=0$.
This corresponds to the $\hat{F}_k$ spanning a set of ``forbidden'' terms for the control Hamiltonian.
One can check this interpretation of the constraint by noticing that if $\hat{U}_t$ solves the Schr\"odinger equation with a Hamiltonian of the form of eqn.(\ref{conprb}),
then variation of eqn.(\ref{eqnfin}) by $\omega_k$ yields:
\begin{align}
\Tr \left(\left(\frac{d \hat{U}_t}{dt} \hat{U}_t^{\dagger} + i\hat{H}_0\right) i\hat{F}_k \right)
\end{align}
which implies:
\begin{align}
\Tr \left(\hat{H}_{c}(t) \hat{F}_k \right) = 0 \nonumber
\end{align}
and thus the desired ``forbidden'' directions are trace-orthogonal to the control Hamiltonian, and the control Hamiltonian has no component in any forbidden direction.
These are essentially identical to the ``linear homogeneous'' constraints in \cite{ACAR2}.
There is a subtle difference however: here the forbidden direction applies only to the control Hamiltonian and not the overall Hamiltonian.
This constraint is equivalent to an \emph{affine} constraint on the overall Hamiltonian.
Adding too many additional constraints may render the system in question uncontrollable.
Existence/uniqueness of optimal trajectories is an issue not addressed in \cite{ACAR2}.

In order to find the equation satisfied by the optimal Hamiltonian that takes into account some additional constraints, we must modify eqn.(\ref{geoeqn1}).
In the remainder of this paper we consider only the ``forbidden direction'' type of additional constraint.
The equations satisfied by the optimal Hamiltonian (if any exist) can be found (in a basis for $\mathfrak{su}(n)$) by variation of each dependent variable on which $\Lambda$ depends.
Variation by $\frac{d \hat{U}_t}{dt}\hat{U}_t^{\dagger}$ yields the EP equation for $\Lambda$:
\begin{align}
\label{sima}
\frac{d}{dt} \frac{\partial \Lambda}{\partial \xi^d} = -C^{a}_{bd} \frac{\partial \Lambda}{\partial \xi^a} \xi^b
\end{align}
Variation by $\omega_k$ yields:
\begin{align}
\label{simb}
 \Tr \left(\left(\frac{d \hat{U}_t}{dt} \hat{U}_t^{\dagger} + i\hat{H}_0\right) i\hat{F}_k \right) = 0
\end{align}
Equations (\ref{sima}) and (\ref{simb}) need to be solved simultaneously in order to obtain the optimal Hamiltonian.

Closed form solutions for these equations and further numerical solution techniques in specific cases of physical interest form the basis of further work.
We intend to perform a complete analysis of common two qubit gates implemented in spin chain systems and other laser driven models.
We hope to obtain an exact formula for the initial conditions required (for the system (\ref{sima}) and (\ref{simb})) in order for numerical solution of the system of ODEs to yield a geodesic connecting $\hat{I}$ to an arbitrary desired gate $\hat{O}$.
A method for achieving a very similar goal has be found in a very different context \cite{imdiff};
only the Riemannian case is addressed, but it seems that the technique is easily adaptable.

\subsection{Role of Sub-Riemannian Geometry and Sub-Finsler Geometry}

Mathematically, the optimal trajectories can be understood as sub-Finsler geodesics,
or more specifically, what could be appropriately called sub-Randers geodesics.
For all the relevant definitions needed here see \cite{ELDSR}.
One can find information about sub-Riemannian geometry in optimal control in \cite{GC}.
For an application of sub-Finsler geometry in optimal control see \cite{SFMOC}.
For a specific application in quantum control see \cite{KN}.
The now well known ``Hormander's condition'' for the controllability of affine linear control systems of the type studied in this work, that is systems of the form of eqn.(\ref{conprb}),
indicates when too many forbidden direction results in the system no longer being controllable.
This would mean that there were unitary gates that could not be implemented using a system constrained in such a way.

The method for handling the constraint $h(i\hat{H}_c(t), i\hat{H}_c(t)) = 1$, alongside additional constraints, shows that the optimal trajectories for $\hat{U}_t$ are geodesics of Randers metric restricted to an affine distribution, say $\mathcal{D}$, on $SU(n)$.
$\mathcal{D}$ is the distribution consisting of vectors in $TSU(n)$ of the form $-i\hat{H}_0\hat{U} + \text{Span}\{ i\hat{H}_k \hat{U} \big| k = 0, \ldots, N \}$.
Here $\{ i\hat{H}_k \big| k = 0, \ldots, N \} \subset \mathfrak{su}(n)$ span the subset of $\mathfrak{su}(n)$ that is $h$-orthogonal to the span of the subset of $\mathfrak{su}(n)$ spanning the ``forbidden directions''.
This distribution is right invariant in the sense that: $\mathcal{D}_{\hat{U}} = \mathcal{D}\hat{U}$.
That is, the optimal trajectories are the length minimising curves that connect given endpoints ($\hat{I}$ to $\hat{O}$) according a Randers metric $F$ (solving the navigation problem in our case), and which are parallel to the distribution $\mathcal{D}$.
A curve $\hat{V}_t$ being parallel to $\mathcal{D}$ means that $\frac{d \hat{V}_t}{dt} \in \mathcal{D}_{\hat{V}_t}$ $\forall t$.
The system is controllable, that is, every unitary gate could be implemented, as long as this distribution is ``Bracket Generating''.
This provides an exact condition, albeit a very technical one, for controlability in the presence of additional constraints.
The systems formed by equations (\ref{sima}) and (\ref{simb}) are solved by such curves.
Eqn.(\ref{sima}) imposes that a curve is an, at least local, extremal curve of the length functional.
Eqn.(\ref{simb}) can then be understood as imposing the curve is parallel according to $\mathcal{D}$.

\subsection{Example Equations For Optimal Trajectories}


We illustrate how a forbidden direction can be treated in the example of a single spin.
For simplicity, we consider the case that there is no drift.
This makes the navigation metric $F$ Riemannian, which makes $\ell$ a quadratic function.
This allows us to solve for the time derivative of $\xi$ explicitly in the EP equations, and then integrate the equation in closed form by hand.
The case with drift is conceptually identical, except it may not always be possible to solve for $\dot{\xi}$, so the resulting system could be more difficult to solve analytically.

We consider the system with control Hamiltonian constrained such that $\frac{1}{2} \Tr(\hat{H}_c(t)^2) = 1$.
Writing $i\hat{H}_c(t) = \xi^k(t) i\sigma_k$ we see that this condition is ${\xi^x}^2 + {\xi^y}^2 + {\xi^z}^2 = 1$.
Suppose further that we are restricted to $\xi^z = c$ for $\forall t$.
This is different from the examples worked out in \cite{ACAR2}, as this is an affine constraint rather than a linear homogeneous constraint.
In a situation with a drift term, it is simple to see that a linear constraint on the control Hamiltonian corresponds to an equivalent affine constraint on the overall Hamiltonian.

In the present case the overall Lagrangian is:
\begin{align}
 \Lambda(\hat{H}_t, \omega(t)) = \frac{1}{2} \left({\xi^x}^2 + {\xi^y}^2 + {\xi^z}^2 \right) + \omega(t) \left(\xi^z - c\right)
\end{align}
The EP equations are:
\begin{align}
\begin{pmatrix}
  \dot{\xi^x} \\ \dot{\xi^y} \\ \dot{\xi^z} + \dot{\omega}
 \end{pmatrix}
 =
 \begin{pmatrix}
  -\omega\xi^y \\ \omega \xi^x \\ 0
 \end{pmatrix}
\end{align}
and $\xi^z(t) = c$, $\dot{\xi^z}=0$,
which implies:
\begin{align}
\begin{pmatrix}
  \dot{\xi^x} \\ \dot{\xi^y}
 \end{pmatrix} =
 \begin{pmatrix}
  -\omega\xi^y \\ \omega \xi^x
 \end{pmatrix}
\end{align}
and that $\omega$ is a constant.
The general solution (after imposing the unit speed condition) is:
\begin{align}
 \xi^x(t) & = A \cos(\omega t) - \sqrt{1-c^2-A^2}\sin(\omega t) \\
 \xi^y(t) & = A \sin(\omega t) + \sqrt{1-c^2-A^2}\cos(\omega t) \nonumber \\
 \xi^z(t) & = c \nonumber
\end{align}
wherein $A$ is an arbitrary constant parameter and $\omega$ is the Lagrange multiplier.
The trajectories in $\mathfrak{su}(n)$ are circles in $\mathfrak{su}(2)$ centered at:
\begin{align}
\begin{pmatrix}
 0 \\ 0 \\ c
\end{pmatrix}
\end{align}
The unit speed condition imposed ensures that the parameter in $\hat{H}_t$ is physical time as it appears in the Schr\"odinger equation.

Both $A$ and $\omega$ paremeterise possible endpoints of $\hat{U}_t$ after the trajectory on the group is reconstructed via the Schr\"odinger equation.
The optimal Hamiltonians are:
\begin{align}
 \hat{H}_t = (A \cos(\omega t) - B\sin(\omega t))\sigma_x + (A \sin(\omega t) + B\cos(\omega t))\sigma_y + c\sigma_z
\end{align}
where $B = \pm \sqrt{1-c^2-A^2}$.

\subsection{Generalising the Forbidden Directions Equations}

The method for handling forbidden directions can be generalised to include a much larger class of constraints replacing the constraint that $h(i\hat{H}_c(t). i\hat{H}_c(t)) = 1$.
One can replace the role of $h$ representing the constraint of the size of $i\hat{H}_c(t)$ with an arbitrary right invariant Finsler metric, which we denote by $\check{F}_{\hat{U}}$, i.e. we now, more generally than before, impose $\check{F}(i\hat{H}_c(t)) = 1$ holds at the identity on $SU(n)$.
As right invariant Finsler metrics on $SU(n)$ are in one-to-one correspondence with Minkowski norms on $\mathfrak{su}(n)$, this new class of constraints is much larger that the class of right invariant Riemannian metrics employed before.

We adapt Shen's \cite{Shen} Lemma 3.1  to the case of $SU(n)$ with a right invariant $\check{F}$.
In addition to the exact solution for the Riemannian case given by Shen's theorem, this gives an equation for a Finlser metric $F$ the geodesics of which are time optimal the the presence of the constraint $\check{F}(i\hat{H}_c(t)) = 1$.
\begin{align}
 \label{navprob}
 \check{F}\left(\frac{\hat{A}}{F(\hat{A})} + i\hat{H}_0 \right) = 1 \qquad \forall \hat{A} \in \mathfrak{su}(n)/\{0\}
\end{align}
Note that the roles of $F$ and $\check{F}$ are reversed here compared with the original presentation.
One can easily check that the solution for $F$ will be a Randers metric exactly when $\check{F}$ is Riemannian; this is exactly the case solved by Shen's theorem.

The premise that the `wind'/drift Hamiltonian can be overcome by the control is now $\check{F}(i\hat{H}_0) \leq 1$. This guarantees that the desired time optimal trajectories are the geodesics of the Finsler metric $F$ solving eqn.(\ref{navprob}).
The solution $F$ is right invariant if both $\check{F}$ and the drift vector field are, as is the case for quantum control problems.
This follows directly from substituting right invariant $\check{F}$ and drift vector field into the equation defining $F$ and then right translating to the identity.

Now we give the set of equation that define the optimal trajectories in such a scenario.
Time optimality yields:
\begin{align}
\frac{d}{dt} \frac{\partial \Lambda}{\partial \xi^d} = -C^{a}_{bd} \frac{\partial \Lambda}{\partial \xi^a} \xi^b
\end{align}
Variation by $\omega_k$ to impose the forbidden direction constraints, as before, yields:
\begin{align}
 \Tr \left(\hat{H}_c(t)\hat{F}_k \right) = 0
\end{align}
Here $\Lambda$ is as before, except the $F$ is no longer necessarily a Randers metric, but is now the solution to eqn.(\ref{navprob}).
This solution is guaranteed to also be a Finsler metric \cite{Shen}.

Together this all yields the system for the time optimal Hamiltonian $\hat{H}_t = \hat{H}_0 + \hat{H}_c(t) = \xi^k \hat{G}_k$:
\begin{align}
\label{toe}
 \begin{dcases}
   \check{F}\left(\frac{\hat{A}}{F(\hat{A})} + i\hat{H}_0 \right) = 1 \ \ \ \ \forall \hat{A} \in \mathfrak{su}(n)/\{0\} \\ \\
   \frac{d}{dt} \frac{\partial \Lambda}{\partial \xi^d} = -C^{a}_{bd} \frac{\partial \Lambda}{\partial \xi^a} \xi^b \\ \\
   \Tr \left(\hat{H}_c(t)\hat{F}_k \right) = 0 \ \ \ \ \forall k, \forall t\geq0 \\ \\
   \hat{U}_T = \mathcal{T} \exp\left(\int_{0}^{T} -i\hat{H}_t dt\right) = \hat{O}
 \end{dcases}
\end{align}
These we refer to as the time optimality equations for the gate $\hat{O}$, the drift Hamiltonian $\hat{H}_0$ and the constraint that $\check{F}(\hat{H}_c(t)) = 1 ~\forall t \geq 0$.
Here, $\hat{G}_k$ are a basis for $\mathfrak{su}(n)$.
These equations determine the optimal Hamiltonian.

As in \cite{ACAR2}, we have not yet found a way to impose the boundary condition $\hat{U}_T = \hat{O}$ (for some $T$) without solving the other time optimality equations explicitly.
It is, however, known which variations at the algebra level correspond to variations of $\hat{U}_t$ that leave the end points of a curve on $SU(n)$ fixed \cite{Hol}.
In quantum mechanical terms these are exactly variations of $-i\hat{H}_t$ of the form: $\delta i\hat{H}_t = i\frac{d \hat{K}_t}{dt} + [i\hat{H}_t, i\hat{K}_t]$.
Here $\hat{K}_t$ is any smooth curve in $\mathfrak{su}(n)$ which is $0$ at both end points.
A method for imposing similar boundary conditions is presented in \cite{imdiff} in a different context.
We hope to analyse that method and adapt it to quantum control scenarios, so that end point conditions on $\hat{U}_t$ can be imposed at the algebra level and thus the EP equations can still be applied.

\section{Conclusions And Further Work}

We have shown that the time optimal control of any quantum system of the form eqn.(\ref{conprb}) with the constraint that $F(i\hat{H}_c(t))=1$, for some Minkowski norm $F$ on $\mathfrak{su}(n)$, is exactly the problem of finding geodesics of a right invariant Finsler metric on $SU(n)$.
Furthermore, we have shown that this can be achieved, for the Hamiltonian driving $\hat{U}_t$ along a geodesic, using the EP equations, which are first order.
We have also shown that, in the presence of forbidden directions for $\hat{H}_c(t)$, the optimal trajectories for $\hat{U}_t$ are the geodesics of a right invariant sub-Finsler metric on $SU(N)$.
Here, sub-Finsler geodesics is taken to mean the shortest curves connecting desired endpoints which are parallel to a specific affine distribution.
We have also shown that this problem can also be expressed as a system of equations in $\mathfrak{su}(n)$ and that these equations are first order also.

We have shown that the method of quantum optimal control based on Randers geometry is highly effective in the case of constraints on the control Hamiltonian of the form $h(i\hat{H}_c,i\hat{H}_c)$ for some inner product $h$ on $\mathfrak{su}(n)$.
We have produced a broadly applicable method that does not depend on the dimension of the Hilbert space that the goal gate $\hat{O}$ acts on.
We believe that our approach will be taken into a more practical setting with further analysis of the ``forbidden directions'' type constraints.
A numerical method for solving eqn.(\ref{sima}) will be presented in further work, along with examples of many practically encountered gates and constraints.

We have also shown that the method can be at least partially generalised to an even broader class of problems where the constraint is represented by a Finsler metric rather than a Riemannian one,
at least in the case that the Finsler metric representing the constraint has the required Killing field, which includes all bi-invariant ones of which there are uncountably many.
One example class of uncountably many bi-invariant Finsler metrics on $SU(n)$ are the formed by the right translation of the Shatten-$p$ norms from the identity.
In this case the desired geodesics can be found in close form.

We intend to produce a general purpose Matlab script into which one can insert:
\begin{itemize}
 \item A drift Hamiltonian
 \item A norm constraining the control Hamiltonian
 \item A desired gate $\hat{O}$
\end{itemize}
and out of which will be produced the optimal control Hamiltonian and control fields, by numerically solving the system of equations (\ref{toe}) and the optimal time.
We predict that the main obstacle to this will be solving for the metric $F$ in terms of $\check{F}$.

Recent work \cite{rage} contains a large appendix ``Euler-Lagrange equation on $SU(n)$'' discussing methods for finding geodesics on $SU(n)$.
Other recent work \cite{NFINCIR} also discusses finding Finsler geodesics on $SU(n)$ in the context of quantum optimal control.
We feel that the relative simplicity of the EP equations, which hold on a vector space $\mathfrak{su}(n)$, compared to the methods described in \cite{rage, NFINCIR} that hold on $SU(n)$, justify the usefulness of our approach.
Furthermore, they avoid the need to ever determine a geodesic on $SU(n)$ when all that is practically needed is the Hamiltonian that drives $\hat{U}_t$ along that geodesic.
Another advantage is the lack of need for the use of any coordinate system on $SU(n)$, as the EP equations directly exploit the right trivisation of $SU(n)$, $TSU(n) \cong SU(n) \times \mathfrak{su}(n)$ (available as Lie groups are all parallelizable manifolds, for definitions see \cite{parman}).

The desired optimal trajectories are geodesics of eqn.(\ref{winmet}).
However, the metric can also be used to obtain optimal times for a system (meeting the appropriate premises) to traverse an \emph{arbitrary curve}.
This is illustrated for the time independent trajectories and a specific drift Hamiltonian and value $h$ in \cite{mememe}.
This is in contrast to other methods which exactly determine optimal trajectories, but do not offer any way to obtain optimal times for arbitrary trajectories.
In practical physical systems, it is unlikely that arbitrary trajectories can be implemented, so one needs a technique for assessing the trajectories that can be implemented over a method for determining theoretically optimal ones.
For example a laser pulse is often described using an envelope function \cite{QOC} and this imposes a form for the control functions a priori.
Our method allows one to assess such systems; a full analysis of a laser driven 2 spin-$\frac{1}{2}$ particle system with a known envelope function will be given in forthcoming work.

Recent work by Brody, Meier and Gibbons \cite{BroB, BroC,BroA} tackles similar navigation problems.
The work in \cite{BroC} specifically obtains identical results to our equations (\ref{eqn:opham}, \ref{optraj}), using an interesting method complementing those in this paper, for the time optimal navigation trajectories and associated control Hamiltonians.

\subsection*{Acknowledgments}

We thank Ian Macintosh for his helpful comments on the Killing fields of invariant metrics on $SU(n)$ and other matters of differential geometry.
We thank Eli Hawkins for his helpful comments on invariant Lagrangians on Lie groups.
We thank Robert Bryant for many helpful comments, answers and references on the MathOverflow Stack Exchange.
We further thank our anonymous reviewers who offered many detailed and helpful comments and corrections.
Russell is supported by an EPSRC studentship.

\bibliographystyle{plain}
\bibliography{p4}

\begin{thebibliography}{10}

\bibitem{TFS}
C.~{Altafini} and F.~{Ticozzi}.
\newblock {Modeling and Control of Quantum Systems: An Introduction}.
\newblock {\em IEEE Transactions on Automatic Control}, 57:1898--1917, 2012.

\bibitem{Fins}
P.~L. Antonelli, editor.
\newblock {\em Handbook of Finsler geometry. vol 1}.
\newblock Kluwer Academic Publishers, 2003.

\bibitem{Bao2004}
David Bao, Colleen Robles, and Zhongmin Shen.
\newblock Zermelo navigation on {R}iemannian manifolds.
\newblock {\em J. Differential Geometry}, 66:377--435, 2004.

\bibitem{imdiff}
Martin Bauer, Martins Bruveris, and Peter~W. Michor.
\newblock Overview of the geometries of shape spaces and diffeomorphism groups.
\newblock {\em Journal of Mathematical Imaging and Vision}, 50(1-2):60--97,
  2014.

\bibitem{RLG}
R.~Berndt.
\newblock {\em Representations of Linear Groups: An Introduction Based on
  Examples from Physics and Number Theory}.
\newblock Vieweg, 2007.

\bibitem{BroB}
Dorje~C. {Brody}, Gary~W. {Gibbons}, and David~M. {Meier}.
\newblock {Time-optimal navigation through quantum wind}.
\newblock {\em New Journal of Physics}, 17, 2015.
\newblock (to appear) \rm arXiv:1410.6724 [quant-ph].

\bibitem{BroA}
Dorje~C. {Brody} and David~M. {Meier}.
\newblock {Elementary solution to the time-independent quantum navigation
  problem}.
\newblock {\em Journal of Physics A}, 48(5):055302, 2015.

\bibitem{BroC}
Dorje~C. {Brody} and David~M. {Meier}.
\newblock {Solution to the quantum Zermelo navigation problem}.
\newblock {\em Physical Review Letters}, 114, 2015.
\newblock (to appear) \rm arXiv:1409.3204 [quant-ph].

\bibitem{Bump}
Daniel Bump.
\newblock {\em Lie Groups}.
\newblock Springer, 2004.

\bibitem{OCQSL}
T.~Caneva, M.~Murphy, T.~Calarco, R.~Fazio, S.~Montangero, V.~Giovannetti, and
  G.~E. Santoro.
\newblock Optimal control at the quantum speed limit.
\newblock {\em Phys. Rev. Lett.}, 103:240501, Dec 2009.

\bibitem{ACAR}
Alberto Carlini, Akio Hosoya, Tatsuhiko Koike, and Yosuke Okudaira.
\newblock Time-optimal quantum evolution.
\newblock {\em Phys. Rev. Lett.}, 96:060503, 2006.

\bibitem{ACAR2}
Alberto Carlini, Akio Hosoya, Tatsuhiko Koike, and Yosuke Okudaira.
\newblock Time-optimal unitary operations.
\newblock {\em Phys. Rev. A}, 75:042308, 2007.

\bibitem{lre}
Hern{\'a}n Cendra, Jerrold~E. Marsden, and Tudor~S. Ratiu.
\newblock Lagrangian reduction by stages.
\newblock {\em Memoirs of the American Mathematical Society}, 152(722), 2001.
\newblock (updated 2009).

\bibitem{ran}
X.~Cheng and Z.~Shen.
\newblock {\em Finsler Geometry: An Approach via Randers Spaces}.
\newblock Springer, 2013.

\bibitem{parman}
L.~Conlon.
\newblock {\em Differentiable Manifolds}.
\newblock Birkh{\"a}user Boston, 2008.

\bibitem{ELDSR}
Enrico~Le Donne.
\newblock Lecture notes on sub-{R}iemannian geometry.
\newblock https://sites.google.com/site/enricoledonne/, 2010.

\bibitem{HOMEGEO2}
Parastoo Habibi, Dariush Latifi, and Megerdich Toomanian.
\newblock Homogeneous geodesics and the critical points of the restricted
  {F}insler function.
\newblock {\em Journal of Contemporary Mathematical Analysis}, 46(1):12--16,
  2011.

\bibitem{hall}
Brian~C. Hall.
\newblock {\em Lie Groups, Lie Algebras, and Representations: An Elementary
  Introduction}.
\newblock Springer, 2003.

\bibitem{Hol}
Darryl~D. Holm, Jerrold~E. Marsden, and Tudor~S. Ratiu.
\newblock The {E}uler--{P}oincar{\'e} equations and semidirect products with
  applications to continuum theories.
\newblock {\em Advances in Mathematics}, 137(1):1--81, 1998.

\bibitem{qconlan}
Michael Hsieh and Herschel Rabitz.
\newblock Optimal control landscape for the generation of unitary
  transformations.
\newblock {\em Phys. Rev. A}, 77:042306, Apr 2008.

\bibitem{FinNav}
Libing Huang and Xiaohuan Mo.
\newblock On geodesics of {F}insler metrics via navigation problem.
\newblock {\em Proc. Amer. Math. Soc.}, 139:3015--3024, 2011.

\bibitem{LAGS}
J.~E. Humphreys.
\newblock {\em Linear Algebraic Groups}.
\newblock Graduate Texts in Mathematics. Springer, 1975.

\bibitem{OTD}
M.~A. {Javaloyes} and M.~{S{\'a}nchez}.
\newblock {On the definition and examples of Finsler metrics}.
\newblock {\em \rm arXiv:1111.5066 [math.DG]}, November 2011.

\bibitem{NewFins}
M.~A. {Javaloyes} and H.~{Vit{\'o}rio}.
\newblock {Zermelo navigation in pseudo-Finsler metrics}.
\newblock {\em \rm arXiv:1412.0465 [math.DG]}, December 2014.

\bibitem{KN}
N.~{Khaneja}, S.~J. {Glaser}, and R.~{Brockett}.
\newblock {Sub-Riemannian geometry and time optimal control of three spin
  systems: Quantum gates and coherence transfer}.
\newblock {\em Phys. Rev. A}, 65(3):032301, March 2002.

\bibitem{LGBAI}
A.~W. Knapp.
\newblock {\em Lie Groups Beyond an Introduction}.
\newblock Birkh{\"a}user Boston, 2002.

\bibitem{EPop}
Wang-Sang Koon and Jerrold~E. Marsden.
\newblock Optimal control for holonomic and nonholonomic mechanical systems
  with symmetry and {L}agrangian reduction, 1995.
\newblock California Institute of Technology, CaltechCDSTR:1995.022.

\bibitem{HOMGEO}
Dariush {Latifi}.
\newblock {Homogeneous geodesics of left invariant Finsler metrics}.
\newblock {\em \rm arXiv:0711.4480 [math.DG]}, 2007.

\bibitem{Dar2}
Dariush Latifi and Megerdich Toomanian.
\newblock Invariant naturally reductive randers metrics on homogeneous spaces.
\newblock {\em Mathematical Sciences}, 6(1), 2012.

\bibitem{Lee}
K.-Y. {Lee} and H.~F. {Chau}.
\newblock {Relation between quantum speed limits and metrics on $U(n)$}.
\newblock {\em Journal of Physics A Mathematical General}, 46(1):015305,
  January 2013.

\bibitem{Li}
B.~{Li}, Z.~{Yu}, S.~{Fei}, and X.~{Li-Jost}.
\newblock {Time optimal quantum control of two-qubit systems}.
\newblock {\em Science China Physics, Mechanics, and Astronomy}, 56:2116--2121,
  2013.

\bibitem{SFMOC}
C.~L\'{o}pez and E.~Mart\'{\i}nez.
\newblock Sub-finslerian metric associated to an optimal control system.
\newblock {\em SIAM J. Control Optim.}, 39(3):798--811, September 2000.

\bibitem{thi}
Andrzej~J. Maciejewski.
\newblock Reduction, relative equilibria and potential in the two rigid bodies
  problem.
\newblock {\em Celestial Mechanics and Dynamical Astronomy}, 63(1):1--28, 1995.

\bibitem{mars}
J.~E. Marsden and T.~S. Ratiu.
\newblock {\em Introduction to Mechanics and Symmetry: A Basic Exposition of
  Classical Mechanical Systems}.
\newblock Springer, 1999.

\bibitem{NFINCIR}
M.~A. {Nielsen}.
\newblock {A geometric approach to quantum circuit lower bounds}.
\newblock {\em Quant.\ Info.\ Comp}, 6:213--262, 2006.

\bibitem{QCOCU}
Jos\'e~P. Palao and Ronnie Kosloff.
\newblock Quantum computing by an optimal control algorithm for unitary
  transformations.
\newblock {\em Phys. Rev. Lett.}, 89:188301, 2002.

\bibitem{plie}
Walter Pfeifer.
\newblock {\em The Lie Algebras $su(N)$: An Introduction}.
\newblock Birkh{\"a}user, 2003.

\bibitem{CRob}
Colleen Robles.
\newblock Geodesics in {R}anders spaces of constant curvature.
\newblock {\em \rm arXiv:math/0501358 [math.DG]}, 2005.

\bibitem{mememe}
Benjamin Russell and Susan Stepney.
\newblock Zermelo navigation and a speed limit to quantum information
  processing.
\newblock {\em Physical Review A}, 90:012303, 2014.

\bibitem{xyzchain}
L.~{\v{S}}amaj and Z.~Bajnok.
\newblock {\em Introduction to the Statistical Physics of Integrable Many-body
  Systems}.
\newblock Cambridge University Press, 2013.

\bibitem{AK}
T.~{Schulte-Herbr{\"u}ggen}, A.~{Sp{\"o}rl}, N.~{Khaneja}, and S.~J. {Glaser}.
\newblock {Optimal control for generating quantum gates in open dissipative
  systems}.
\newblock {\em Journal of Physics B Atomic Molecular Physics}, 44(15):154013,
  2011.

\bibitem{Shen}
Z.~{Shen}.
\newblock {Finsler Metrics with K = 0 and S = 0}.
\newblock {\em {\rm arXiv:math/0109060 [math.DG]}}, September 2001.

\bibitem{GC}
G.~Stefani.
\newblock {\em Geometric Control Theory and Sub-Riemannian Geometry}.
\newblock Springer, 2014.

\bibitem{mlog}
J.~Stillwell.
\newblock {\em Naive Lie Theory}.
\newblock Springer, 2008.

\bibitem{rage}
X.~{Wang}, M.~{Allegra}, K.~{Jacobs}, S.~{Lloyd}, C.~{Lupo}, and M.~{Mohseni}.
\newblock {Quantum brachistochrone curves as geodesics: obtaining accurate
  control protocols for time-optimal quantum gates}.
\newblock {\em \rm arXiv:1408.2465 [quant-ph]}, 2014.

\bibitem{QOC}
J.~{Werschnik} and E.~K.~U. {Gross}.
\newblock {Quantum Optimal Control Theory}.
\newblock {\em \rm arXiv:0707.1883 [quant-ph]}, 2007.

\bibitem{Yosh}
Ryozo Yoshikawa and SorinV. Sabau.
\newblock Kropina metrics and {Z}ermelo navigation on {R}iemannian manifolds.
\newblock {\em Geometriae Dedicata}, 171(1):119--148, 2014.

\end{thebibliography}

\end{document}